\documentclass[aps,prl,twocolumn,superscriptaddress]{revtex4-1}
\usepackage{graphicx}
\usepackage{subfigure}
\usepackage{epstopdf}
\usepackage{amsmath}
\usepackage{amssymb} 
\usepackage{amsfonts}
\usepackage{mathrsfs}
\usepackage{theorem}
\usepackage{bm}
\usepackage{url}
\usepackage{csquotes}
\usepackage{algorithm}
\usepackage{algorithmicx}
\usepackage{algpseudocode}
\usepackage{dcolumn}
\usepackage{color}
\usepackage{natbib}

\begin{document}
	
	\title{Quantum autoencoders using mixed reference states}
	

\author{Hailan Ma}
\affiliation{School of Engineering and Technology, University of New South Wales, Canberra, ACT 2600, Australia}
\affiliation{CIICADA Lab, School of  Engineering, The Australian National University, Canberra, ACT 2601, Australia}
\author{Gary J. Mooney}
\affiliation{School of Physics, University of Melbourne, Parkville, Victoria 3010, Australia}
\author{Ian R. Petersen}
\affiliation{CIICADA Lab, School of  Engineering, The Australian National University, Canberra, ACT 2601, Australia}
\author{Lloyd C. L. Hollenberg} 
\affiliation{School of Physics, University of Melbourne, Parkville, Victoria 3010, Australia}
\author{Daoyi Dong}
\email{daoyi.dong@anu.edu.au}
\affiliation{CIICADA Lab, School of  Engineering, The Australian National University, Canberra, ACT 2601, Australia}
\affiliation{School of Physics, University of Melbourne, Parkville, Victoria 3010, Australia}
\date{\today}

\begin{abstract}
	One of the fundamental tasks in quantum information theory is quantum data compression, which can be realized via quantum autoencoders that first compress quantum states to low-dimensional ones and then recover to the original ones with a reference state. When taking a pure reference state, there exists an upper bound for the encoding fidelity, which limits the compression of states with high entropy. To overcome the entropy inconsistency, we allow the reference state to be a mixed state and propose a cost function that combines the encoding fidelity and the quantum mutual information.  We consider the reference states to be a mixture of maximally mixed states and pure states and propose three strategies for setting the ratio of mixedness. Numerical simulations of different states and experimental implementations on IBM quantum devices illustrate the effectiveness of our approach. 
	
\end{abstract}

\maketitle

\section{Introduction}\label{sec:introduction}

Quantum machine learning which combines machine learning and quantum computation has grown into a booming research topic~\cite{biamonte2017quantum,dong2008quantum,huang2021power,cerezo2022challenges,niu2019universal,li2020quantum,dong2022quantum}. Quantum autoencoders (QAEs) inherit the spirit of classical autoencoders to compress information into a latent space such that the original data can be recovered from a reduced-dimension representation~\cite{pu2016variational,bartuuvskova2006optical}. They have the potential to reduce the requirements of quantum communication channels~\cite{steinbrecher2019quantum} and the size of quantum gates~\cite{lamata2018quantum,ding2019experimental} and thus have a practical value for various applications including quantum simulation~\cite{aspuru2005simulated}, quantum communication and distributed computation in quantum networks~\cite{steinbrecher2019quantum,lamata2018quantum}. 


Owing to the potential of QAEs in quantum information processing, there is a growing interest in designing different schemes to complete state compression tasks. An early work proposed a quantum generalization of a classical neural network~\cite{wan2017quantum} and another work designed an autoencoder framework using programmable circuits~\cite{romero2017quantum}.  An enhanced QAE that encodes the feature vector of the input data into single-qubit rotation gates has been implemented in variational quantum circuits~\cite{bravo2021quantum}. There have also been achievements in the implementation of QAEs on photonic systems~\cite{pepper2019experimental,huang2020realization,ding2019experimental}. Apart from data compression, QAEs have also been applied to other applications, such as state denoise~\cite{bondarenko2020quantum,achache2020denoising} and error mitigation~\cite{zhang2021generic}. A novel method based on QAEs has been devised to prepare the quantum Gibbs state and estimate the quantum Fisher information ~\cite{du2021exploring}. A hybrid QAE has been proposed to identify the emergence of order in the latent space that can be utilized for clustering and semi-supervised classification~\cite{srikumar2021clustering}. Recently, the execution of a QAE-facilitated teleportation protocol has been implemented on a silicon photonic chip~\cite{zhang2022resource}. Furthermore, the exploration of QAEs to analyze datasets originating from industrial contexts demonstrates their potential in processing real-world data~\cite{mangini2022quantum}.

In traditional QAE schemes~\cite{romero2017quantum,pepper2019experimental,huang2020realization,ma2023compression}, pure states are utilized as reference states for recovering the initial state. For each state to be compressed, there exists an upper bound (hereafter, we call it QAE-pure bound) for the encoding fidelity, i.e., the overlap between the trash state and the reference state. Such a bound limits the compression of states with high entropy.  To compensate for the entropy inconsistency between the initial state and the recovered state~\cite{cao2021noise}, we allow the reference state to be a mixed one. Instead of a fixed mixed state, we configure different reference states for effective compression of different states. To achieve this, we take advantage of quantum mutual information that measures the disentanglement to guide the optimization of the encoding transformation. Meanwhile, the conventional cost function in standard QAEs aims to decouple the initial states into two parts and provides information about the inner structure of initial states, which can be useful for setting the reference state. Hence, we design a novel cost function that combines the above two factors to guide the training of QAEs towards better performance. Inspired by the compression of tensor products of identical states on IBM quantum devices~\cite{pivoluska2022implementation}, we experimentally realize QAEs with mixed reference states on the IBM quantum simulator \emph{ibmq$\_$qasm$\_$simulator} and quantum computer \emph{ibmq$\_$quito}. 



In this work, we leverage mixed reference states to break the upper bound of compression rate imposed by conventional QAEs~\cite{ma2023compression}. In particular, we use a mixture of a pure state and the maximally mixed state, with pure reference states being a special case. Similarly, our proposed cost function considers both quantum mutual information and the existing function in conventional QAEs that favour pure reference states. Consequently, our protocol exhibits flexibility, enabling its application to various quantum states.  We have observed the direct relationship between the optimal value of the purity ratio in the reference state and the QAE-pure bound. This discovery empowers us with the insights gained from the training of QAEs to adaptively configure mixed reference states for compressing different unknown states. Our experimental implementations on IBM quantum computers demonstrate the potential of the proposed protocol in saving valuable quantum resources for real applications.


\section{Results}\label{sec:results}

\begin{figure}
	\centering
	\centerline{\includegraphics[width=3.4in]{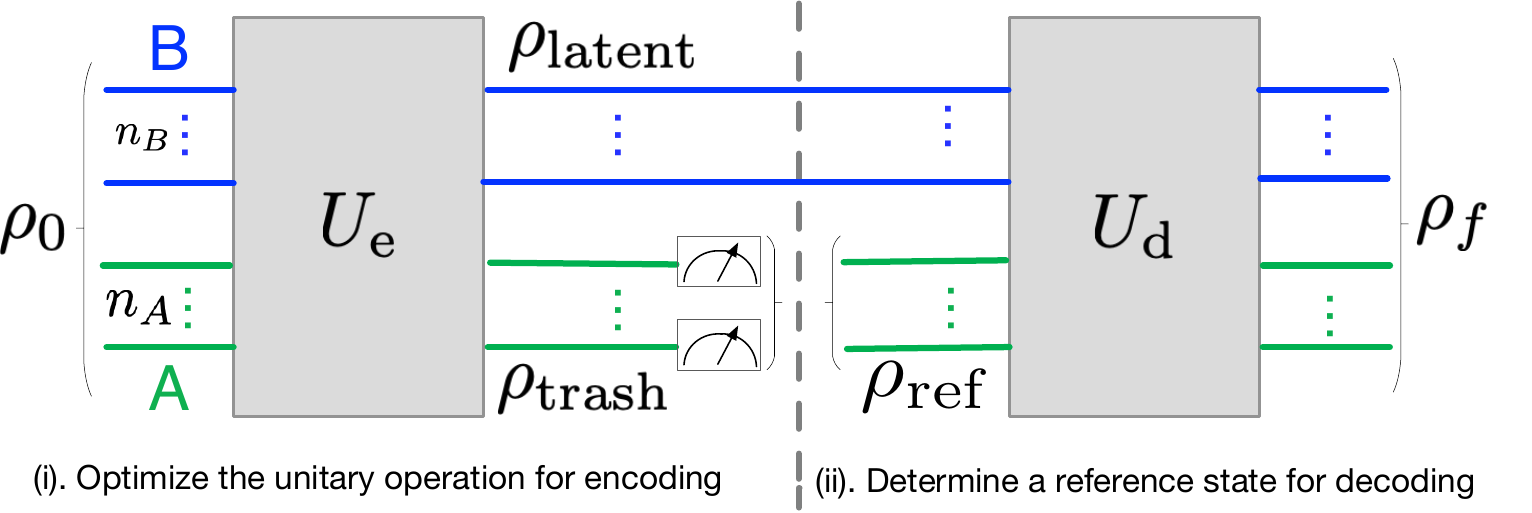}}
	\caption{ Quantum circuit representation of a QAE using a mixed reference state. The network includes two parts: (i) The encoder $U_{\rm e} $ reorganizes the $(n_A+n_B)$-qubit initial state $\rho_0$ into two parts, i.e., $n_B$-qubit state $\rho_{\rm latent}$ contains the useful information (blue lines) that represent the latent qubits and $n_A$-qubit state $\rho_{\rm{trash}}$ contains the superfluous information (green lines) that represent the trash qubits. Here, the trash state is obtained by tracing out the latent space of $n_B$ qubits.  The latent state is obtained by tracing out the trash space of $n_A$ qubits. (ii) The decoder $U_{\rm d} $ recovers the state $\rho_f$ by using the combination of the latent state and ancillary fresh qubits (initialized to the reference state). The goal of QAEs is to maximize the overlap between the recovered state $\rho_f$ and the original state $\rho_0$. Note that, we use the common practice of taking the decoding as the inverse of the encoding of $U_{\rm d} =U_{\rm e} ^{\dagger}$~\cite{pepper2019experimental,ma2023compression}. }
	\label{fig:autoencoder}
\end{figure}

\subsection{QAEs using mixed reference states}
\textbf{Schematic.} As illustrated in Fig.~\ref{fig:autoencoder}, we define the trash qubits as subsystem $A$ and the latent qubits as subsystem $B$, respectively. We denote the dimensions of the original space, the latent space, and the trash space as $d$, $d_B$, and $d_A$, respectively. The goal of a QAE is to compress $(n_A+n_B)$-qubit state $\rho_0$ into $n_B$-qubit state $\rho_{\rm latent}$ via an encoder map $U_{\rm e} $ and then recover to $(n_A+n_B)$-qubit state $\rho_f$ via a decoder map $U_{d}$.  After the encoding operation $U_{\rm e} $, the trash state and the latent state are obtained as $\rho_{\rm{trash}} = \textup{Tr}_{B}(U_{\rm e} \rho_0 U_{\rm e} ^{\dagger})$ and $\rho_{\rm latent} = \textup{Tr}_{A}(U_{\rm e} \rho_0 U_{\rm e} ^{\dagger})$, respectively. Denote $F(\rho_1,\rho_2)$ as the state fidelity between $\rho_1$ and $\rho_2$~\cite{nielsen2010quantum}.
The efficiency of this task can be quantified by the decoding fidelity between the original state and the recovered state, i.e., $\mathcal{F}_{\rm d}=F(\rho_0,\rho_f)$ and the scheme is considered reliable when $\mathcal{F}_{\rm d}$ approaches $1$. During the whole process, a reference state is utilized for two aspects: (i) measure the encoding fidelity between the trash state and the reference state, denoted as $\mathcal{F}_{\rm e}=F(\rho_{\rm{trash}},\rho_{\rm{ref}})$; (ii) reproduce the initial states with the combination of the latent state and the reference state. When the unitary operation $U_{\rm e} $ perfectly disentangles $\rho_0$ into two parts as 
$U_{\rm e}  \rho_0 U_{\rm e} ^{\dagger}=\rho_{\rm latent} \otimes \rho_{\rm{ref}}  $, the overlap between the trash state and the reference state can achieve unity~\cite{romero2017quantum,ma2023compression} and the decoding fidelity $\mathcal{F}_{\rm d}$ can also achieve unity. In this work, the unitary transformation $U_{\rm e} $ is realized through Hamitonian-based control (see Section~\textbf{Methods} for detailed information).

When taking a pure state as reference states $\rho_{\rm{ref}}=|\psi_{\rm ref}\rangle \langle \psi_{\rm ref}|$ for compressing the initial state $\rho_0$. There exists an upper bound (i.e., QAE-pure bound, abbreviated as $Q_{\rm pure}^{\rm bound}$) for the encoding fidelity between  $\rho_{\rm{trash}}$ and $\rho_{\rm{ref}}$. From the previous work~\cite{ma2023compression}, we have
	\begin{equation}
		Q_{\rm pure}^{\rm bound}(\rho_0) = \max_{U_{\rm e} } F(\textup{Tr}_{A}(U_{\rm e} \rho_0 U_{\rm e} ^{\dagger}), |\psi_{\rm ref}\rangle) = \sum_{k=1}^{d_B} \lambda_k(\rho_0),
		\label{eq:qaebound}
	\end{equation}
	where $\lambda_k(\rho)$ is the $k$-th (in descending order) eigenvalue of $\rho$. This bound is determined by eigenvalues of the initial state $\rho_0$,  with no dependency on the pure reference state $|\psi_{\rm ref}\rangle$~\cite{ma2023compression}. Hence, in traditional QAEs, a common practice is to utilize a fixed pure state as the reference state, e.g., $|0\rangle \langle 0|$. According to Eq.~(\ref{eq:qaebound}),  if $\rho_0$ has a rank larger than  $d_B$, the optimal encoder $U_{\rm e} $ can only decouple the largest $d_B$ eigenvalues of $\rho_0$, whose sum is less than one~\cite{ma2023compression,cao2021noise}. As such, a high-rank state in this work means the rank of its density matrix is larger than $d_B$. 

When compressing $\rho_0$ with high entropy, the trash state $\rho_{\rm{trash}}$ tends to have high entropy and consequentially have low overlap with a pure reference state (e.g., setting $\rho_{\rm{ref}}=|0\rangle \langle 0|$).  In the decoding stage, the low entropy of a pure state may also limit the entropy of the recovered state (see $\rho_f$ in Fig.~\ref{fig:autoencoder}). To overcome the entropy inconsistency between the initial state $\rho_0$ and the recovered state $\rho_f$~\cite{cao2021noise}, we remove the limitation of a pure reference state and allow the reference state $\rho_{\rm{ref}}$ to be mixed. The limitation of the conventional QAEs also motivates us to adopt different mixed states rather than a fixed state for compressing different initial states. In this way, the entropy in the reference state can assist the decoder in achieving a high fidelity for the recovered state. Here, the introduction of entropy offers an intuitive strategy for setting mixed reference states. To ensure that the recovered state $\rho_f$ has high fidelity with $\rho_0$, additional efforts are required to optimize $U_{\rm e} $. Instead of searching $U_{\rm e} $ and $\rho_{\rm{ref}}$ together (a full encoding and decoding procedure is required), we accomplish the task of QAEs with mixed reference states within two stages.

\textbf{Cost function.} In conventional QAEs, the reference state is fixed as a pure state, and $\mathcal{F}_{\rm e}$ is utilized as the cost function to train QAEs~\cite{romero2017quantum,pepper2019experimental,huang2020realization,ma2023compression}. However, $\mathcal{F}_{\rm e}$ is different from $\mathcal{F}_{\rm d}$ which characterizes the effectiveness of QAEs in compressing and recovering quantum data. When allowing the reference state to be a mixed state, $\mathcal{F}_{\rm e}$ can achieve one by setting $\rho_{\rm{ref}}=\rho_{\rm{trash}}$, whereas $\mathcal{F}_{\rm d}$ is usually less than one and can reach one only when perfect disentanglement is realized~\cite{romero2017quantum}. Given that quantum mutual information (QMI) measures the correlation between subsystems of quantum states~\cite{wilde2011classical,watrous2018theory}, it quantifies the amount of noise that is required to erase (destroy) the correlations completely. To facilitate QAEs using mixed reference states, we aim to disentangle $ U_{\rm e}  \rho_0 U_{\rm e} ^{\dagger}$, which is achieved by minimizing the QMI, i.e.,  $\mathcal{I}(U_{\rm e}  \rho_0 U_{\rm e} ^{\dagger})$ or  equivalently maximizing 
\begin{equation}
	J_{\rm{qmi}}=- \mathcal{I}(U_{\rm e}  \rho_{0} U_{\rm e} ^{\dagger}),
\end{equation}
where $
\mathcal{I}(\rho)=S(\textup{Tr}_A(\rho))+S(\textup{Tr}_B(\rho))-S(\rho)$ denotes the QMI of $\rho$ and $S(\rho)=-\textup{Tr}(\rho\ln(\rho))$ denotes the von Neumann entropy of $\rho$. Generally $\mathcal{I}(\rho) \geq 0$, and $\mathcal{I}(\rho)=0$ when $\rho=\textup{Tr}_A(\rho) \otimes \textup{Tr}_B(\rho)$. 

According to existing research~\cite{ma2023compression}, when training QAEs using the overlap between the trash state and a pure reference state (e.g., taking the reference state as $\rho_{\rm ref}=|0\rangle \langle 0|$), 
$J_{\rm{pure}}=F(\textup{Tr}_B(U_{\rm e}  \rho_0 U_{\rm e} ^{\dagger}),|0\rangle  \langle 0|) $ as the cost function, the compression rate of QAEs can approach the theoretical QAE-pure bound. Under that scheme, a high compression rate can be realized for low-rank states~\cite{ma2023compression}. Although this cost function fails to disentangle a high-rank state with satisfactory performance, the optimization of $J_{\rm{pure}}$ leads to a direction of reorganizing the information of initial states into two parts. To combine the cases of low-rank states and high-rank states, we propose a cost function,
\begin{equation}
	\Phi({\rm{w}})= {\rm{w}}J_{\rm{pure}}+(1-{\rm{w}})J_{\rm{qmi}},
	\label{eq:unifiedloss}
\end{equation}
where $\rm w\in[0,1]$ controls the ratio of different factors. This protocol is termed QAE-qmi in this paper. Considering the potential of evolutionary strategy (ES) in conventional QAEs~\cite{ma2023compression}, we employ it to optimize the parameters of $U_{\rm e} $.

\textbf{Reference states.} 
Recall the nature of QAEs lies in disentangling~\cite{romero2017quantum}. The encoding fidelity that measures the overlap between the trash states and the reference states can reach one (i.e., $\mathcal{F}_{\rm e}=1$) by setting $\rho_{\rm{ref}}=\rho_{\rm{trash}}$. Although $\mathcal{F}_{\rm d} \leq \mathcal{F}_{\rm e}$, in the general case, $\mathcal{F}_{\rm d}$ can approach $\mathcal{F}_{\rm e}$ and they both achieve one when perfect disentangling is realized~\cite{ma2023compression}. When there is no limitation for the reference state, it is helpful to investigate the performance of QAE-qmi with $\rho_{\rm{ref}}=\rho_{\rm{trash}}$. 

In practical applications, it may be useful to utilize reference states with some physical constraints. According to our previous study, a pure reference state, e.g., $|0\rangle \langle 0|$ is effective in compressing low-rank states, with the compression rate approaching the QAE-pure bound~\cite{ma2023compression}, whose value is usually high for low-rank states. For high-rank states with high entropy, the introduction of mixed reference states helps increase the entropy of the recovered states~\cite{cao2021noise}. While the maximally mixed states $I /{d_A}$ has the highest entropy among all states in $\mathcal{H}_{A}$ and is effective for increasing the entropy in the decoding stage. To achieve a good QAE for different quantum states, we take the following reference state
\begin{equation}
	\rho_{\rm mix}=p_r |0\rangle  \langle 0| +(1-p_r) I /  d_A,
	\label{eq:mixedref}
\end{equation}
where $p_r$ represents the ratio of the pure state and $(1-p_r)$ represents the ratio of the mixed state in the reference state. $I$ denotes the $d_A$-dimensional identity matrix. Different initial states with different inner structures may have different optimal reference states following the form of Eq.~(\ref{eq:mixedref}). When compressing initial states with high entropy, it is preferable to use low $p_r$ that generates high entropy for $\rho_{\rm mix}$. As such, it is desirable to specify an optimal $p_r$ for different quantum states.  Although mixed reference states cost additional memories, our method aims to achieve high fidelity between the recovered state and the initial state. This is particularly important for initial states with low QAE-pure bound. By constraining the mixed reference state in the form of Eq.~(\ref{eq:mixedref}), one can transmit the compressed latent representation and $p_r$ to facilitate the subsequent recovery.

Now, we focus on determining a good $p_r$ to recover quantum states with high decoding fidelity. Intuitively, quantum states with different inner structures (entropy) require different optimal $p_r$ to achieve optimal decoding fidelity. Before deciding the optimal $p_r$ for recovering the state, we first propose a grid-search strategy for setting $p_r$ (marked as \emph{grid}). Under a fixed cost function (e.g., $\Phi=0.5(J_{\rm{pure}}+J_{\rm{qmi}})$), we define a  candidate set (e.g., $\{0,0.1,0.2,0.3,0.4,0.5,0.6,0.7,0.8,0.9,1.0\}$) for $p_r$. Then,  following the optimized $U_{\rm e} $ during the encoding process, the decoding process is performed with different $p_r$ selected from the candidate pool. Eventually, the optimal one is determined regarding its decoding fidelity, and this value is defined as $p_r^{\rm grid}$. In the following, we also introduce another two strategies: (i) \emph{bound}: to leverage the QAE-pure bound when prior knowledge of the initial state is available; (ii) \emph{guess}: to infer from the training process of QAEs. The \emph{guess} strategy acts as a practical solution that can be implemented in different cases.

In summary, the optimization of QAEs using a mixed reference state is accomplished within two stages. Firstly, the training of QAEs using $\Phi({\rm {w}})={\rm{w}} J_{\rm{pure}}+(1-{\rm w})J_{\rm{qmi}}$ is implemented in the encoding stage, while the mixed reference state is introduced in the decoding stage. After the optimization of $U_{\rm e} $ is finished by maximizing $\Phi$, we determine a reference state for recovery to maximize the overlap between the initial state and the recovered state. 

\subsection{Numerical results}

\textbf{Quantum state settings.}
For the compression task, we consider three classes of quantum states.  Firstly, we consider thermal states as 
\begin{equation}
	\rho=\frac{e^{-\beta H}}{\textup{Tr}(e^{-\beta H})}, 
\end{equation}
where $\beta$ is the inverse temperature and $H$ denotes the Hamitonian. Let $\sigma_z^j $($\sigma_x^j$) denote the composite value of $\sigma_z$($\sigma_x$) on the $j$-th qubit with identity matrices for the other qubits. For example, we investigate thermal states with the Hamiltonian of the one-dimensional transverse-field 
	\begin{equation}
		H=-(\sum_j \sigma_z^j\sigma_z^{j+1}+\sum_j \sigma_x^j),
	\end{equation}
	with couplings set to 1. Then, we investigate Werner states, which are bipartite and are invariant under any unitary operator in the form of $U\otimes U$~\cite{lyons2012werner}. Let $|k\rangle$ and $|j\rangle$ be the computational basis for two bipartite subspaces, respectively.  A Werner state can be parameterized by 
\begin{equation}
	\rho(\alpha)=\frac{1}{d^2-d\alpha}(I-\alpha \sum_{kj}|kj\rangle \langle jk|),
\end{equation}
where $I$ denotes the $d^2$-dimensional identity matrix and $\alpha$ varies between -1 and 1. Additionally, we also consider the initial states that have a similar form to $\rho_{\rm mix}$ as 
\begin{equation}
	\rho(p_0)= p_0 |\psi \rangle \langle \psi|+(1-p_0) I / d,
	\label{eq:mixedinitial}
\end{equation} 
where $I$ denotes the $d$-dimensional identity matrix and the value of $p_0$ controls the purity of the initial states. In this work, we randomly generate a pure state $|\psi\rangle$ and utilize it for different values of $p_0$. By computation, we find that states with high $p_0$ have a high QAE-pure bound. In this work, we consider compressing 2-qubit states into 1-qubit states and 4-qubit states into 2-qubit states.  The density matrices of 2-qubit thermal states and Werner states are presented in  Supplementary Discussion 1. 

\begin{figure}
	\centering
	\includegraphics[width=3.5in]{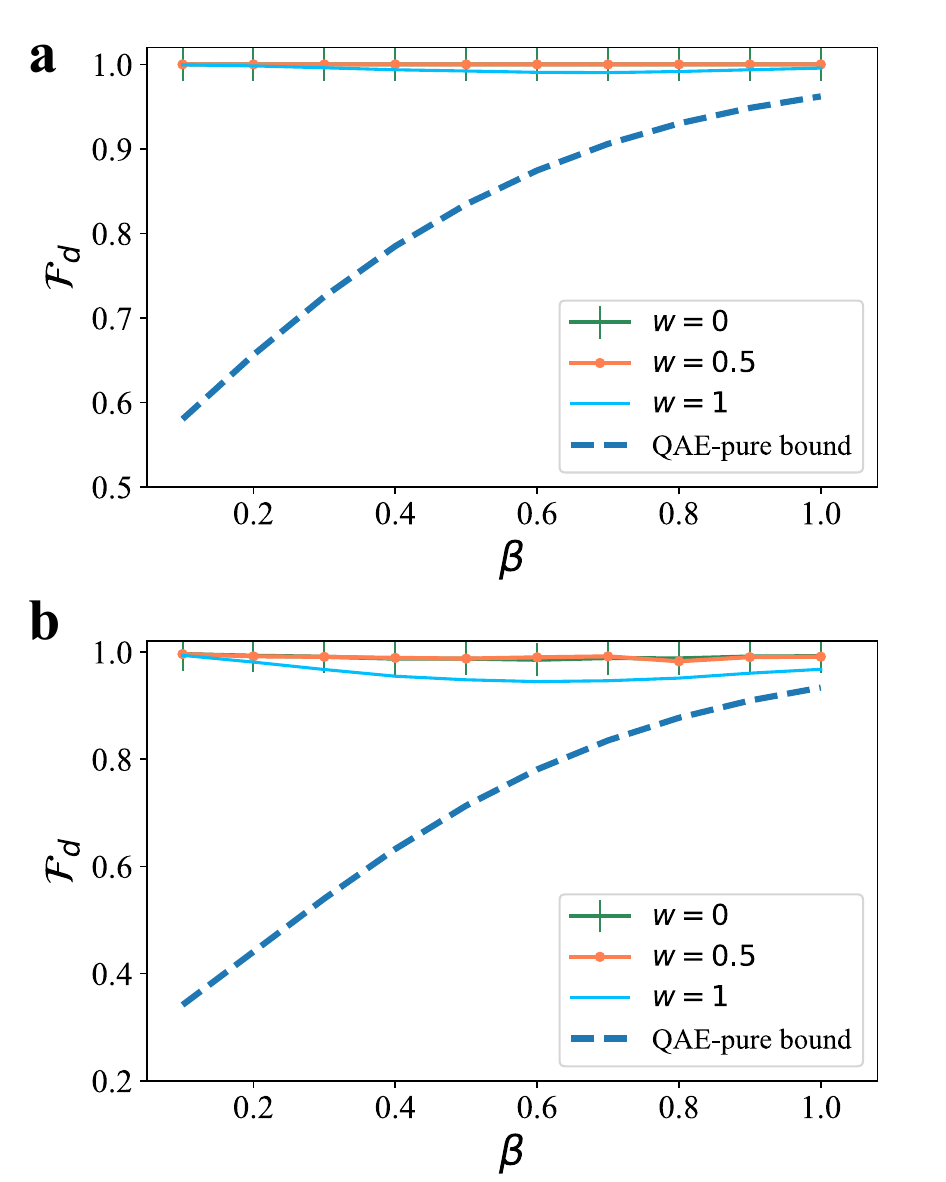}
	\caption{Comparison of QAE-qmi under different $\rm w$ using $\rho_{\rm{ref}}=\rho_{\rm{trash}}$. \textbf{a} for 2-qubit states, \textbf{b} for 4-qubit states. $\beta$ controls the entropy of the thermal states, i.e., an increase of $\beta$ leads to a decrease of entropy in quantum states.  $\mathcal{F}_{\rm d}$ represents the decoding fidelity between the initial state and the recovered state. $w$ denotes the ratio of $J_{\rm pure}$ in the cost function $\Phi(\rm w)$. The blue dashed line represents the theoretical upper bound of compression rate when only considering pure reference states~\cite{ma2023compression}. }
	\label{fig:Jab-thermal}
\end{figure}

\begin{figure}
	\centering
	\includegraphics[width=3.5in]{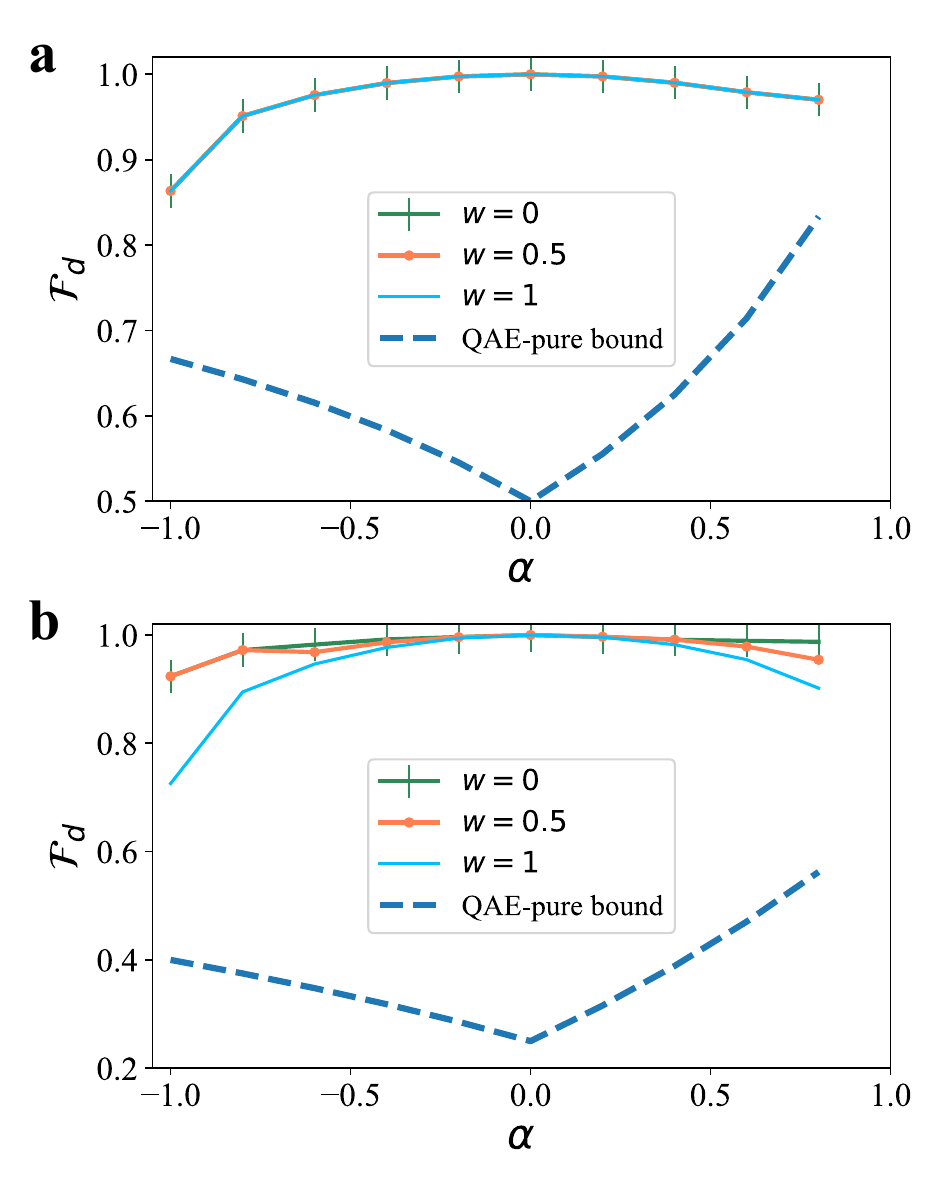}
	\caption{Comparison of QAE-qmi under different $\rm w$ for Werner states under $\rho_{\rm{ref}}=\rho_{\rm{trash}}$.  \textbf{a} for 2-qubit states, \textbf{b} for 4-qubit states. $\rm w$ denotes the ratio of $J_{\rm pure}$ in the cost function $\Phi(\rm w)$. $\mathcal{F}_{\rm d}$ denotes the fidelity between the initial state and the recovered state. The blue dashed line represents the theoretical upper bound of compression rate when only considering pure reference states~\cite{ma2023compression}.}
	\label{fig:Jab-werner}
\end{figure}

\textbf{Investigation of different $\rm w$.} Firstly, we investigate the performance of $\rho_{\rm ref} = \rho_{\rm trash}$. In particular, we consider three different cases:  (i) $\rm w=1$ considers the fidelity between the trash state and a fixed pure state; (ii) $\rm w=0$ considers QMI of the encoding state $U_{\rm e}  \rho_0 U_{\rm e} ^{\dagger}$; (iii) $\rm w=0.5$ considers both factors and acts as a more general function. Comparsion results for 2/4-qubit thermal states and 2/4-qubit Werner states are provided in Fig.~\ref{fig:Jab-thermal}  and Fig.~\ref{fig:Jab-werner}, respectively. These results demonstrate that $\Phi=J_{\rm{qmi}}$ and $\Phi=0.5(J_{\rm{pure}}+J_{\rm{qmi}})$ achieve a  similar decoding fidelity, higher than that of $\Phi=J_{\rm{pure}}$.  The gaps between $\rm w=0.5$ and $\rm w=1$ suggest that introducing quantum mutual information to the cost function $\Phi(\rm w)$ helps enhance the decoding fidelity for 4-qubit states. These results suggest that the introduction of QMI is effective in compressing and recovering quantum states, especially for states with low QAE-pure bounds (i.e., thermal states with $\beta$ approaching 0 and Werner states with $|\alpha|$ approaching 0). The training curves of QAE-qmi under different $\rm w$ are summarized in Supplementary Discussion 2.  

Then we turn to the case of setting reference states as Eq.~(\ref{eq:mixedref}). Although the disentanglement evaluation of $J_{\rm qmi}$ naturally follows the nature of QAEs, this function fails to achieve a perfect value of 1 for non-separable initial states. When considering $\rho_{\rm ref}=\rho_{\rm mix}$ for low-rank states, the optimization of $J_{\rm{qmi}}$ tends to bring the trash state far away from a pure state, in conflict with the intuition that pure reference states are enough for low-rank states. To combine the two scenarios together,  $\rm{w} =  0.5$ is a good solution. 

\begin{figure*}
	\centering
	\includegraphics[width=7.0in]{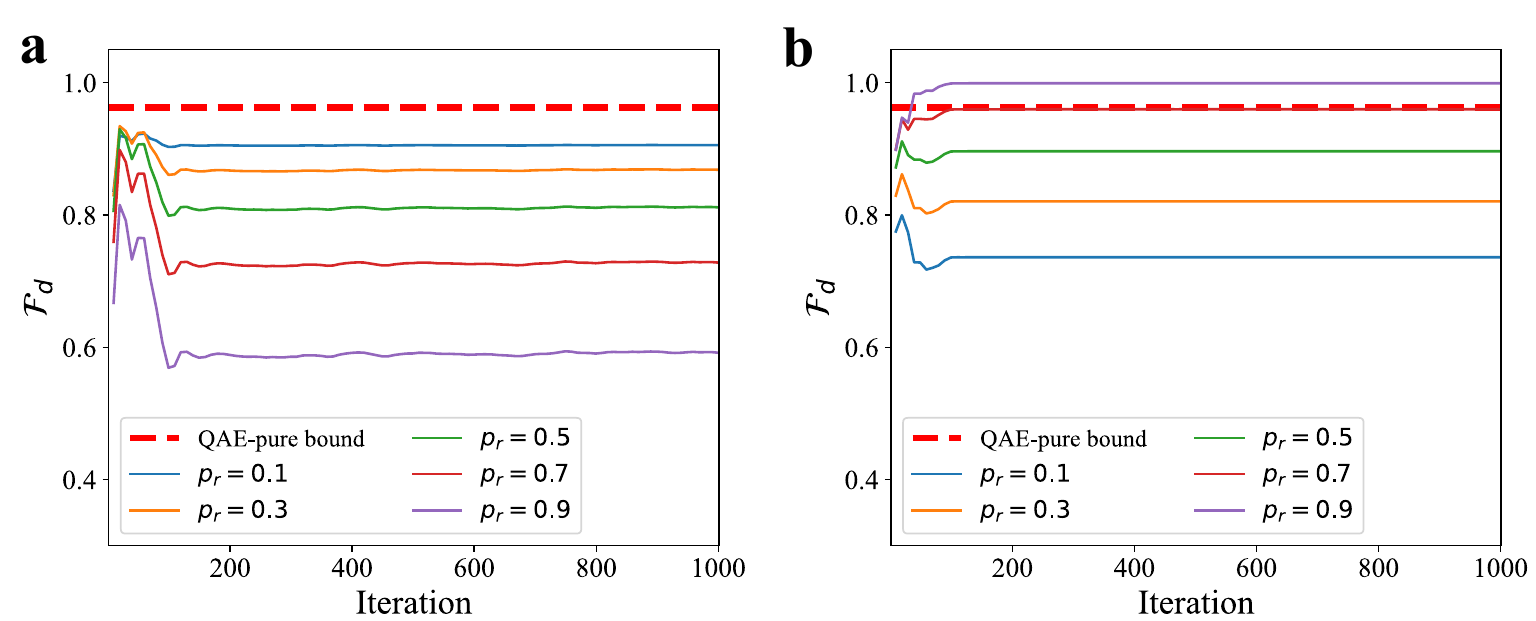}
	\caption{An example of QAE-qmi with $\rho_{\rm ref}=\rho_{\rm mix}$ under different $p_r$ for 2-qubit thermal states.  \textbf{a} for $\rm w =0$, \textbf{b} for $\rm{w}=0.5$. $p_r$ represents the ratio of the pure state in the reference state $\rho_{\rm mix}$ and $(1-p_r)$ represents the ratio of the mixed state in the reference state $\rho_{\rm mix}$. $\mathcal{F}_{\rm d}$ represents the decoding fidelity between the initial and recovered states. The red dashed line represents the theoretical upper bound of compression rate when only considering pure reference states~\cite{ma2023compression}. In this figure, some selected values of $p_r$ from the candidate pool are used to clearly demonstrate the trends of different values.}
	\label{fig:failure}
\end{figure*}

We implement an example of low-rank states (with high QAE-pure bound) using $\rho_{\rm{ref}}=\rho_{\rm mix}$ under different values of $p_r$, with results in Fig.~\ref{fig:failure}. While, the best solution under $\rm w=0$ corresponds to a large $p_r$ (indicating high entropy in $\rho_{\rm mix}$) and  the best solution under $\rm w=0.5$ corresponds to a small $p_r$ (indicating low entropy in $\rho_{\rm mix}$). The latter case is in line with our conjecture that compressing states with low entropy and high QAE-pure bound requires a reference state with high purity. As such, we consider $\Phi=0.5(J_{\rm{pure}}+J_{\rm{qmi}})$  to be useful for both $\rho_{\rm{ref}}=\rho_{\rm{trash}}$ and $\rho_{\rm{ref}}=\rho_{\rm mix}$. Hereafter, without specific notation. QAE-qmi refers to the case of $\Phi=0.5(J_{\rm{pure}}+J_{\rm{qmi}})$.

The cost function in Eq.~(\ref{eq:unifiedloss}) can be applied to compress initial states with different purities. According to Eq.~(\ref{eq:qaebound}), states with high purity tend to have eigenvalues close to 0 or 1, and have a high QAE-pure bound~\cite{ma2023compression}, suggesting that the conventional QAE approach (with only $J_{\rm pure}$) still works. In the unified approach, it is preferable to adopt a large $\rm w$ for compressing initial states with high purities (e.g., thermal states with large  $\beta$ or Werner states with $|\alpha|$ approaching 1). Furthermore, when compressing initial states with high purity, for example, $\rho(p_0)$ with high $p_0$, the existence of QMI in the cost function with ratio $\rm w=0.5$ brings in a negative effect, which can be overcome by reducing the ratio of $J_{\rm qmi}$ e.g., $\rm w=0.99$. This originates from the two stages of our protocol: firstly optimize $U_{\rm e} $ via $\Phi(\rm w)$ and then determine a reference state for recovering. A large ratio of $J_{\rm qmi}$ tends to conflict with the mechanism of setting reference states as $p_r |0\rangle \langle 0|+(1-p_r)I/d_A$. This conflict increases with the purity of states, as initial states with high purities favour a high ratio of $J_{\rm pure}$ and a high value of $p_r$. Nevertheless, we can avoid this by adjusting $\rm w$. Please refer to Supplementary Discussion 4 for detailed information.

\begin{figure}
	\centering
	\includegraphics[width=3.5in]{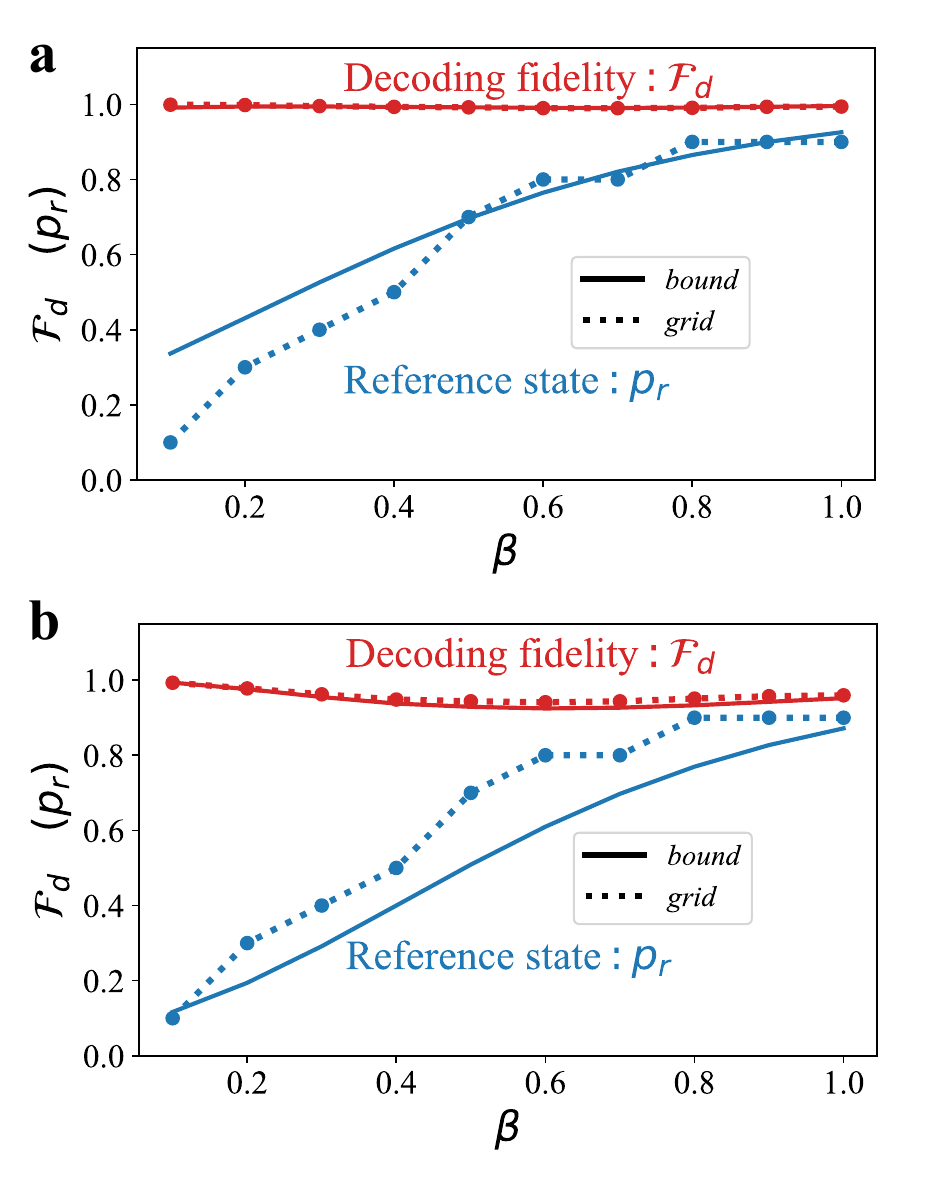}
	\caption{Comparison of two strategies of setting $p_r$ for compressing and recovering thermal states.  \textbf{a} for 2-qubit states, \textbf{b} for 4-qubit states. $p_r$ represents the ratio of the pure state and $(1-p_r)$ represents the ratio of the mixed state in the reference state. $\mathcal{F}_{\rm d}$ represents the fidelity between the initial state and the recovered state. The solid blue line (dashed blue line) corresponds to the actual value of $p_r$. The solid red line (dashed red line) represents the value of $\mathcal{F}_{\rm d}$ through the bound information (grid searching from a set of values).}
	\label{fig:fitting-thermal}
\end{figure}

\begin{figure}
	\centering
	\includegraphics[width=3.5in]{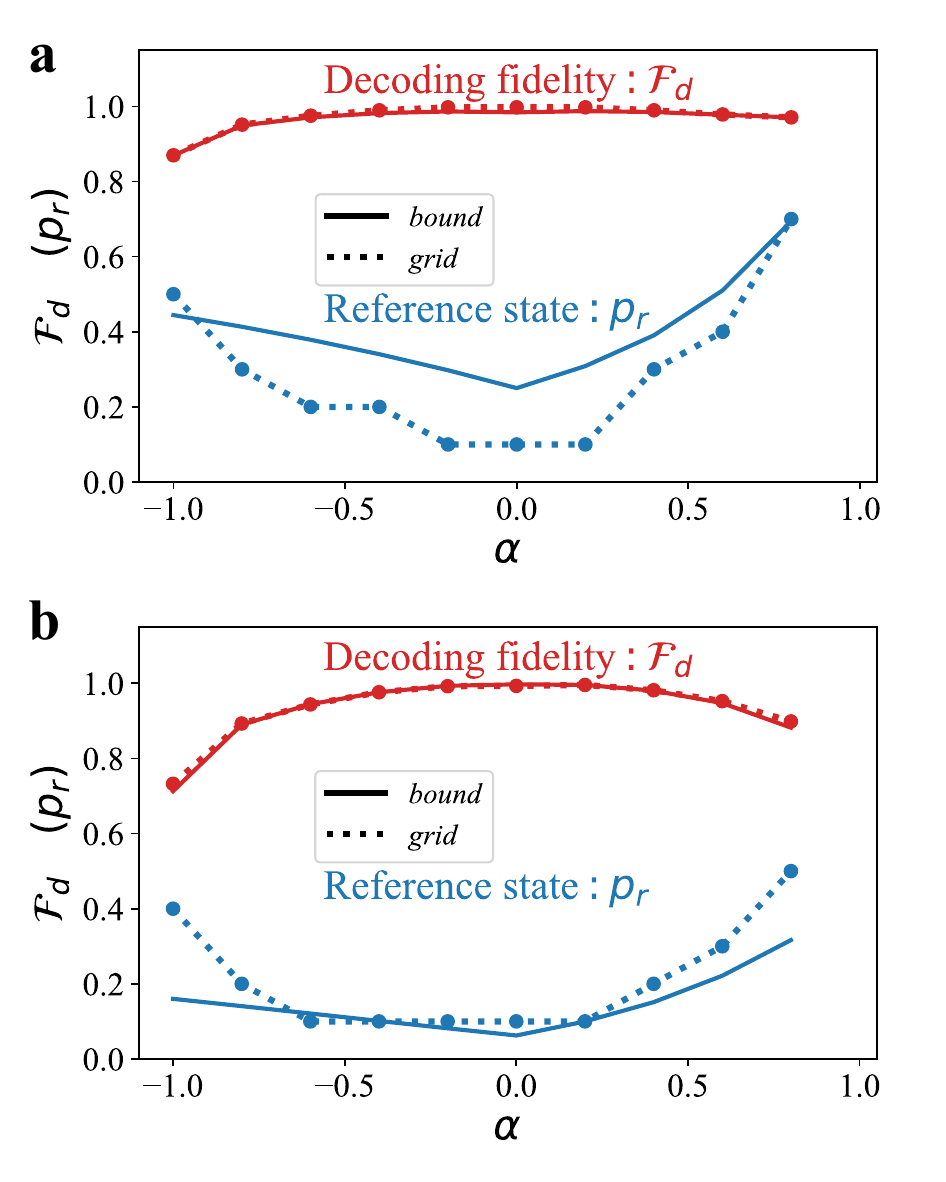}
	\caption{Comparison results of two strategies of setting $p_r$ when compressing and recovering Werner states.  \textbf{a} for 2-qubit states, \textbf{b} for 4-qubit states. $\mathcal{F}_{\rm d}$ represents the fidelity between the initial and recovered states. $p_r$ represents the ratio of the pure state and $(1-p_r)$ represents the ratio of the mixed state in the reference state. The solid blue line (dashed blue line) corresponds to the actual value of $p_r$. The solid red line (dashed red line) represents the value of $\mathcal{F}_{\rm d}$ through the bound information (grid searching from a set of values).}
	\label{fig:fitting-werner}
\end{figure}

\begin{figure}
	\centering
	\includegraphics[width=3.5in]{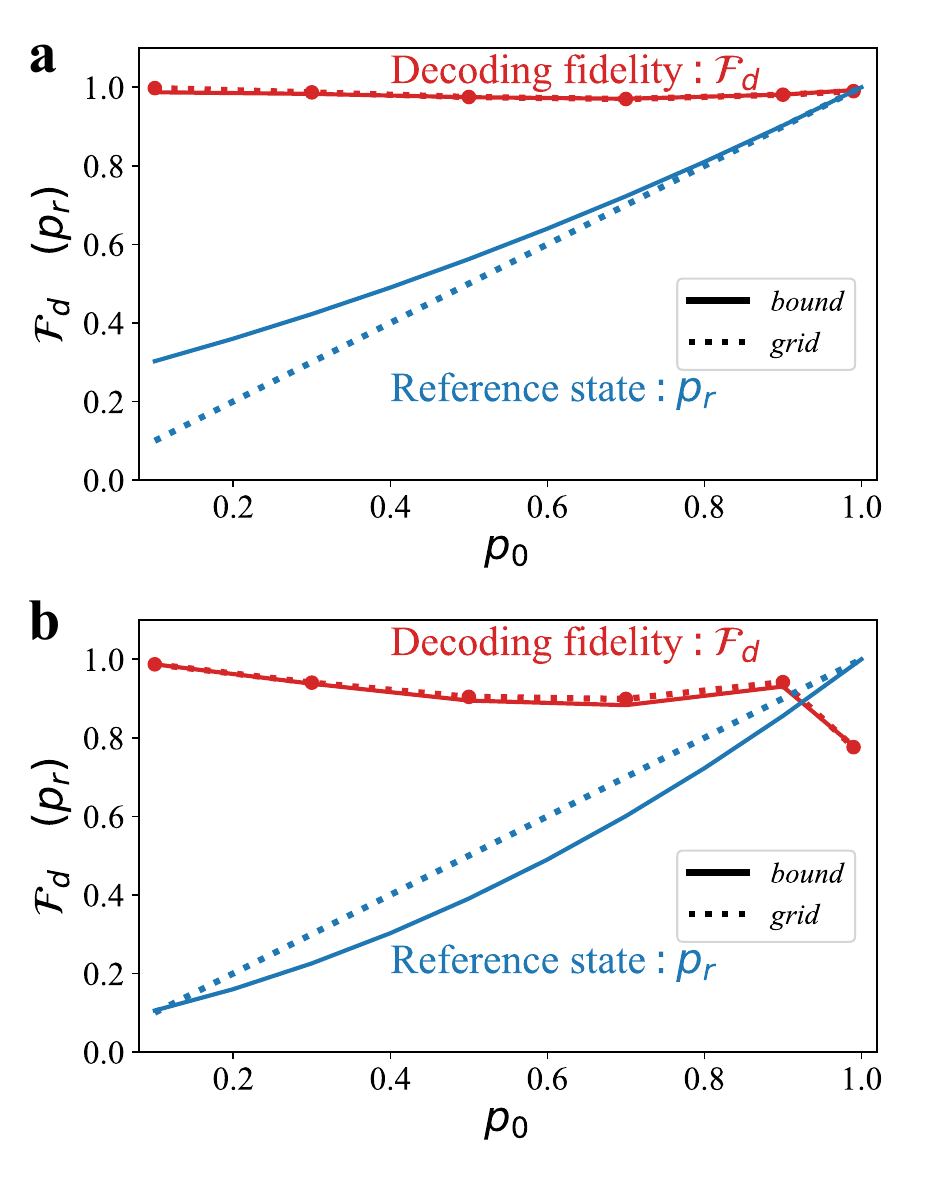}
	\caption{Comparison of two strategies of setting $p_r$ for compressing maximally mixed states blended with pure states.  \textbf{a} for 2-qubit states, \textbf{b} for 4-qubit states. $p_r$ represents the ratio of the pure state and $(1-p_r)$ represents the ratio of the mixed state in the reference state. $\mathcal{F}_{\rm d}$ represents the fidelity between the initial state and the recovered state. The solid blue line (dashed blue line) corresponds to the actual value of $p_r$. The solid red line (dashed red line) represents the value of $\mathcal{F}_{\rm d}$ through the bound information (grid searching from a set of values). }
	\label{fig:fitting_haar}
\end{figure}

\textbf{Investigation of different $p_r$.} The introduction of mixed states aims to bring in appropriate entropy for recovering different states. Intuitively, high-rank states with high entropy and low QAE-pure bounds require more entropy for recovery, i.e., a low $p_r$. Recall that each state has its inner structure, and can be characterized by a QAE-pure bound~\cite{ma2023compression}. Initially, we tested $Q_{\text{pure}}^{\text{bound}}$ but found that its value does not align well with $p_r^{\text{grid}}$, and their performance regarding $\mathcal{F}d$ exhibits a significant gap. Subsequently, we find the square of $Q_{\text{pure}}^{\text{bound}}$ aligns more closely with $p_r^{\text{grid}}$, and their decoding fidelities are comparably close. Hence, we propose a second strategy (marked as \emph{bound}) for setting $p_r$ by leveraging the QAE-pure bound of the initial state $\rho_0$, and we have $p_r^{\rm bound}=(Q_{\rm pure} ^{\rm bound}(\rho_0))^2$. 

To validate our conjecture that the optimal $p_r^{\rm grid}$ tends to approach the square of the QAE-pure bound, we compare the two strategies, with their actual values of $p_r$ and the associated decoding fidelity $\mathcal{F}_{\rm d}$ in one figure. From the results in Fig.~\ref{fig:fitting-thermal}, Fig.~\ref{fig:fitting-werner}, it is clear that $p_r^{\rm grid}$ has the same trend as $p_r^{\rm bound}$, and their decoding fidelities are close to each other under different parameters.  For states in the form of $\rho(p_0)$, we observe that the best $p_r$ found using the \emph{grid} strategy (i.e., $p_r^{\rm grid}$) tends to be approaching $p_0$. The comparison of the two strategies (\emph{grid} vs \emph{bound}) regrading $p_r$ and $\mathcal{F}_{\rm d}$ is summarized in  Fig.~\ref{fig:fitting_haar}. The two curves are close to each other with increasing $p_0$. The decoding fidelities for the two strategies of setting $p_r$ achieve similar values. 

Based on the observation that taking $p_r$ as the square of QAE-pure bound achieves good performance, $J_{\rm{pure}}$ among the unified cost function is approaching the QAE-pure bound during the training process~\cite{ma2023compression}. We propose a third strategy (marked as \emph{guess}) for adaptively configuring the reference states as $p_r^{\rm guess}=J_{\rm{pure}}^2$. The comparison results of different strategies of setting $p_r$ are summarized in Supplementary Discussion 3, demonstrating that the manual and automatic ways of setting $p_r$ are effective in compressing and recovering different quantum states. By now, we have three strategies for setting $p_r$, and their comparsion for compressing thermal states and Werner states is summarized in Supplementary Discussion 3. Under the strategy of $p_r=p_r^{\rm guess}$, we further compare the performance of QAE-qmi under different $\rm w$, with results shown in Supplementary Discussion 4. The decline of $\mathcal{F}_{\rm d}$ when $p_0$ approaches one reveals that  $\Phi=0.5(J_{\rm{pure}}+J_{\rm{qmi}})$ hinders the compression of quantum states with high $p_0$.  

\subsection{Experimental results}

\begin{figure*}
	\centering
	\centerline{\includegraphics[width=7in]{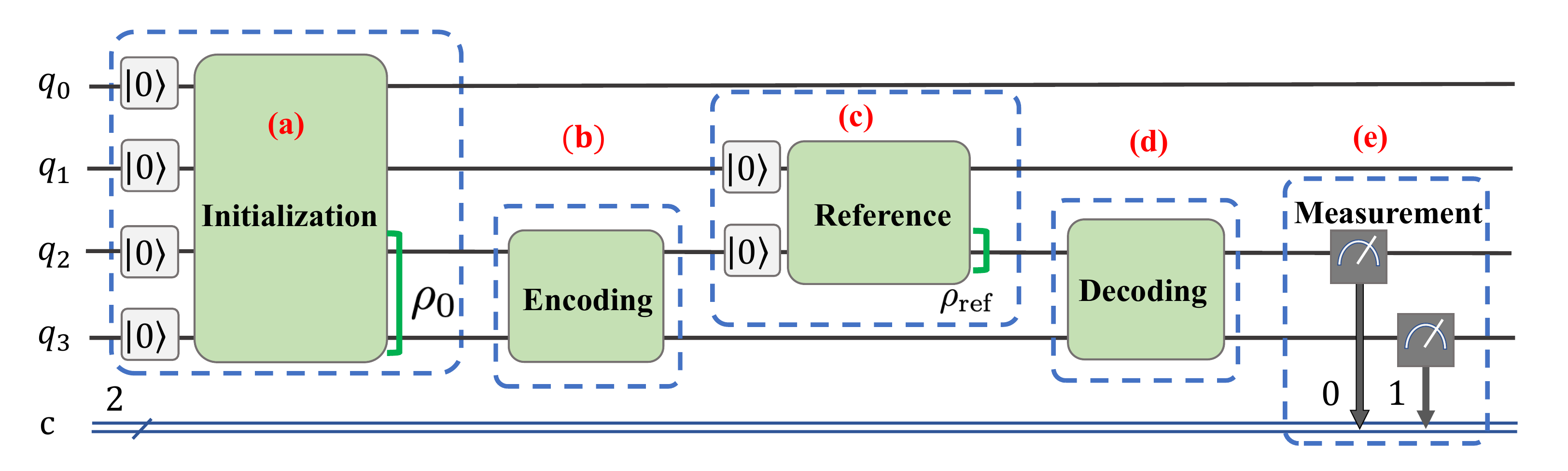}}
	\caption{Quantum circuits for compressing 2-qubit states into 1-qubit states and recovering to 2-qubit states using mixed reference states. Firstly, four qubits are required to generate a 4-qubit pure state in (a), whose partial trace on $q_0q_1$ leads to a mixed state $\rho_0$ on $q_2q_3$.  Hereafter, the encoding operation and the decoding operation are performed on $q_2q_3$, where $q_2$ contains the trash state  $\rho_{\rm{trash}}$ and $q_3$ contains the latent state $\rho_{\rm latent}$. Then, $q_1q_2$ are reused to generate a pure state whose partial trace on $q_1$ leads to a reference state $\rho_{\rm{ref}}$ on $q_2$ in (c). Then, the decoding operation in (d) is performed on the combination of $q_2$ with $\rho_{\rm{ref}}$ and $q_3$ with $\rho_{\rm latent}$. In (e), measurements are performed on $q_2q_3$ for quantum state tomography to obtain the density matrix of the recovered state.} 
	\label{fig:ibmcircuit}
\end{figure*}

\begin{figure*}
	\centering
	\includegraphics[width=7.0in]{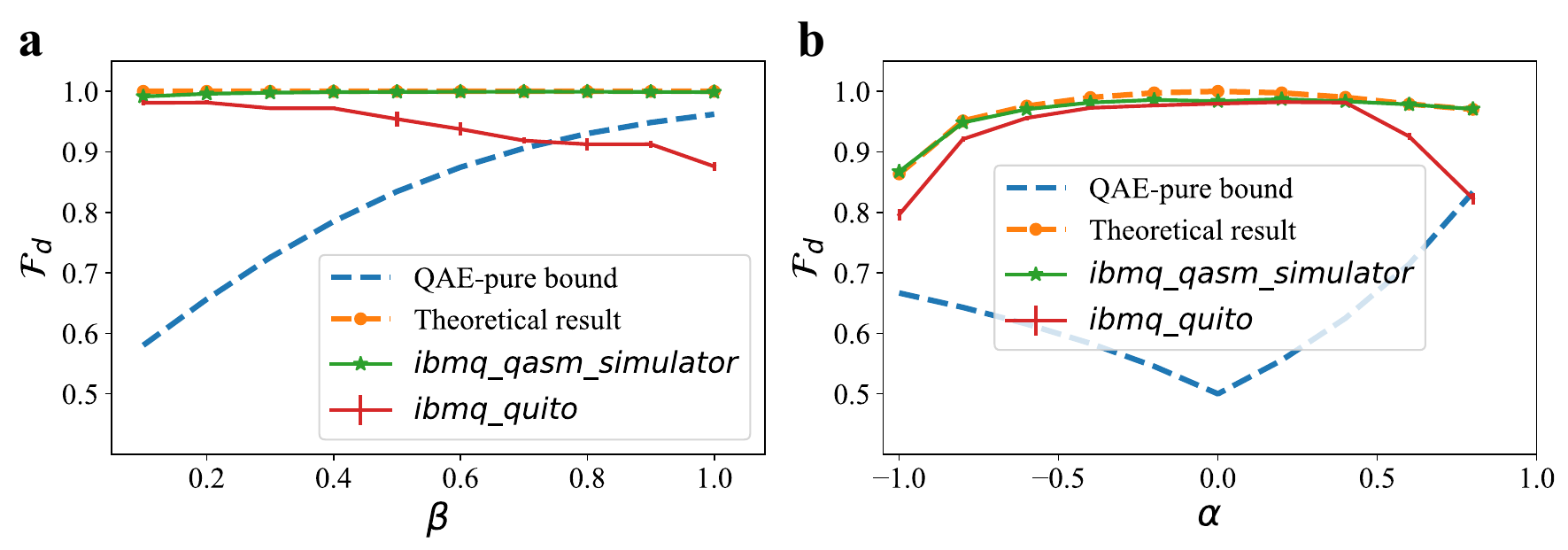}
	\caption{Results for QAEs on \emph{ibmq$\_$qasm$\_$simulator} and \emph{ibmq$\_$quito} for encoding and decoding 2-qubit states.  \textbf{a} for thermal  states, \textbf{b} for Werner states. $\mathcal{F}_{\rm d}$ represents the fidelity between the initial state and the recovered state.  The blue dashed line represents the upper bound of compression rate when only considering pure reference states~\cite{ma2023compression}. The orange line represents the theoretical results of simulating QAEs on classical computers. The green line represents the numerical results obtained from the IBM simulator \emph{ibmq$\_$qasm$\_$simulator}. The red line with error bars represents the average results of 6 times together with standard deviation when quantum circuits are sampled with 8192 shots on IBM Quantum device \emph{ibmq$\_$quito}. }
	\label{fig:ibmresults}
\end{figure*}

Generally, it is assumed that quantum circuits deal with pure states. We need to find a solution to generate mixed states, which is essential in preparing initial states and reference states for encoding and decoding, respectively. We use the technique of \emph{purification}~\cite{nielsen2010quantum} to associate a mixed state with a pure state in a large space. Given a state $\rho_K$ of a quantum system $K$, it is possible to introduce another system $R$, and define a pure state $|\psi\rangle_{KR}$ for the joint system $KR$ such that $\rho_K=\textup{Tr}_R(|\psi\rangle \langle \psi|_{KR})$. The pure state $|\psi\rangle_{KR}$ reduces to $\rho_K$ when we look at the system $K$ alone. This mathematical procedure can be done for any state. Please refer to Supplementary Discussion 5 for detailed information about the construction of $|\psi\rangle_{KR}$ for arbitrary $\rho_K$. 

The quantum circuit for compressing 2-qubit states into 1-qubit states is depicted in Fig.~\ref{fig:ibmcircuit}, where four qubits  $q_0q_1q_2q_3$ are utilized to generate mixed states on $q_2q_3$, on which the encoding gate and the decoding gate are performed. The circuit can be divided into five parts: (a) prepare the initial state, (b) perform the encoding operation, (c) prepare the reference state, (d) perform the decoding operation, (e) perform quantum measurements to obtain the density matrix of the recovered state. A set of complete measurements is required to specify the density matrix of a quantum state~\cite{dong2022quantum}.  In Fig.~\ref{fig:ibmcircuit}, only a special case of local measurement of $\sigma_z\otimes \sigma_z$ on $q_2q_3$ is performed.  Adding some gates (such as the Hadamard gate and the $S^{\dagger}$ gate) before the measurement part (i.e., between (d) and (e)) helps realize other measurements. Hence, the quantum circuits are repeated several times until a complete measurement is accomplished. Feeding the measured data to the built-in function for \emph{quantum state tomography} in qiskit~\cite{ibmQST2021}, the density matrix of the recovered state is finally obtained.  

In this work,  we do not perform the optimization loops on quantum devices. Instead, we take the encoding transformation $U_{\rm e} $ and the reference state in the form of Eq.~(\ref{eq:mixedref}) that are learned numerically on classical computers, and then deploy them on IBM quantum simulators and quantum computers, respectively.  Note that, each green block in Fig.~\ref{fig:ibmcircuit} represents quantum circuits composed of a sequence of quantum gates to achieve unitary operations. Please refer to Supplementary Discussion 5 for the transpiled circuits for the green blocks. 

We implement the procedure of compressing and recovering 2-qubit states on \emph{ibmq$\_$qasm$\_$simulator} and  \emph{ibmq$\_$quito}, with $8192$ shots. Each compression task is run 6 times on \emph{ibmq$\_$quito}. The comparison results are summarized in Fig.~\ref{fig:ibmresults}. The results of the simulators are in agreement with the theoretical results obtained from classical computers. However, gaps exist between the results of \emph{ibmq$\_$qasm$\_$simulator} and \emph{ibmq$\_$quito}. In particular, the gap becomes apparent for thermal states with increasing $\beta$ and Werner states with  $\alpha$ approaching -1 or 1. The underlying reason may be that compressing states close to maximally mixed states with a low QAE-pure bound, presents a large space for improvement through the introduction of mixed reference states. By contrast, initial states closer to pure states can achieve a high compression rate by utilizing pure reference states, while pure states can be affected by various noise sources, e.g., CNOT noise. Then, we visualize the density matrices of the initial states and the recovered states obtained from ~\emph{ibmq$\_$quito} for Werner states in Supplementary Discussion 5. 

\section{Discussion}\label{section:discussion}

In this paper, we have investigated the performance of QAEs with mixed reference states.  One may consider employing a fixed mixed state as the reference state as in conventional QAEs. However, it is challenging to decide on a fixed reference state for different initial states, and the proof that arbitrary mixed reference states yield similar bounds remains elusive. The way of merging all the operations into the encoder fails to reveal features of QAEs with mixed reference states. By comparsion, the adaptive configuration of mixed reference states provides a clear clue about improving fidelity via appropriate entropy compensations. Then, we summarize the characteristics of our protocol as follows.

(i) The proposed function of $\Phi({\rm w})={\rm w} J_{\rm{pure}}+(1-{\rm w}) J_{\rm{qmi}}$ combines the approximate QAE-pure bound function that reflects the inner structure of the initial states and the quantum mutual information that measures the correlation between subsystems. It is a general function that can be applied to both low-rank states and high-rank states. As demonstrated by the numerical results, training QAEs using $\Phi(\rm w)$ achieves high decoding fidelity under different reference setting rules including  $\rho_{\rm{ref}}=\rho_{\rm{trash}}$ and $\rho_{\rm{ref}}=\rho_{\rm mix}$. In addition, it has been found that for initial states with high QAE-pure bounds (e.g., large $p_0$ in Eq.~(\ref{eq:mixedinitial})), it is preferable to increase $\rm{w}$ in $\Phi({\rm w})={\rm w} J_{\rm{pure}}+(1-{\rm w}) J_{\rm{qmi}}$, giving more importance to the approximate QAE-pure bound. This is consistent with the fact that pure reference states together with $J_{\rm{pure}}$ can realize a good compression rate for low-rank states~\cite{ma2023compression}. However, it is crucial to recognize the inherent limitation of employing QMI, whose value depends on the von Neumann entropy that is not an observable. Consequently, this limitation restricts its applications in experimental settings. In our future work, we will explore an approximate function that can be more readily implemented in experimental contexts.

(ii) The numerical results demonstrate that setting the reference state in the form of Eq.~(\ref{eq:mixedref}) helps enhance the decoding fidelity for high-rank states. Due to the special form of the reference states, it is intuitive that different initial states may rely on different optimal purity ratios $p_r$ that help maintain the entropy consistency between the initial states and the recovered states. As demonstrated by the numerical results in Fig.~\ref{fig:fitting-thermal} and Fig.~\ref{fig:fitting-werner}, and Fig.~\ref{fig:fitting_haar}, the optimal $p_r$ using via the \emph{grid} strategy is close to the square of QAE-pure bound for thermal states, Werner states and maximally mixed states blended with pure states. Such findings provide hints for adaptively setting reference states for different quantum states. 

(iii) When limiting the reference states to the form of Eq.~(\ref{eq:mixedref}), we can take advantage of the prior information of the initial states to determine a mixed reference state that achieves a high $\mathcal{F}_{\rm d}$. For example, $(Q_{\rm pure}^{\rm bound}(\rho_0))^2$ can be determined before the training process of QAEs. If no prior knowledge is available, we can also approximate the QAE-pure bound by inferring from the training process and taking $F(\rho_{\rm trash},|0\rangle)^2$. From this perspective, our protocol may have wide applications in practical quantum applications.  For example,  the compressed latent representations can be utilized to effectively denoise errors in the original states~\cite{bondarenko2020quantum}, or act as intermediate states to facilitate high-dimensional subspace teleportation~\cite{zhang2022resource}.

Our work illustrates the effectiveness of QAEs using mixed reference states under different constraints and thus provides implications for practical applications. More work remains to be investigated in the future. For example, other forms of mixed reference states are worthy of further exploration. Imperfections in quantum system models are not considered in this work. Our future work will also include general quantum channels to deal with decoherence for mixed quantum states.

\section{Methods}\label{section:methods}

\subsection{Quantum control model}\label{subsec:model}


Here, we use the density matrix $\rho(t)$ (which is a Hermitian, positive semidefinite matrix satisfying $\textup{Tr}(\rho(t))=1$ to describe the state of a closed quantum system. The evolution equation for $\rho(t)$ can be described by the quantum Liouville equation~\cite{dong2010quantum}
\begin{equation}
	{\rm{i}}\dot{\rho}(t)=\rho(t)H(t)-H(t)\rho(t).
	\label{eq:liou}
\end{equation}
When we use control fields $\{u_j(t)\}_{j=1}^{M}$ to manipulate the system, the system Hamiltonian in Eq.~(\ref{eq:liou}) can be divided into two parts, i.e., $H(t)=H_0+H_c(t)=H_0+\sum_{j=1}^{M}u_j(t)H_j,$ where $H_0$ is the time-independent free Hamiltonian of the system, $H_c(t)$ is the control Hamiltonian representing the interaction of the system with the control fields. For such a control system, its solution is given as $\rho(t)=U (t)\rho_0 U^{\dagger}(t)$ with $U(0)=I$, where the propagator $U(t)$ is formulated as follows:
\begin{equation}
	\frac{d}{dt}U(t)=-{\rm{i}} (H_0+\sum_{j=1}^{M}u_j(t)H_j)U(t).
	\label{eq: unitary propagator}
\end{equation}

For the compression task, we consider spin chain models with  $$H_0=\sum_{i=0}^{n_A+n_B-1}\sigma_i^{x}\sigma_{i+1}^{x}+\sigma_i^{y}\sigma_{i+1}^{y}+\sigma_i^{z}\sigma_{i+1}^{z}.$$ Chains with Heisenberg coupling are known to be controllable given at least two noncommuting controls acting on the first or the last spin in the chain~\cite{burgarth2009local,wang2016subspace}, we exert control fields on the first two qubits towards $X$ and $Y$ directions~\cite{dong2023learning}, with the control Hamiltonian as
$$H_c=\sum_{k=0,1} u_k^x\sigma_k^x+u_k^y \sigma_k^y.$$
As such, there are four control fields to be designed. We use piece-wise control fields, which means that the total control time $T$ is equally divided into different periods, with each having $\rm{dt}=T/N$ duration times. In this work, the total control time $T=20$ is equally divided into 100 pieces. The bound of control fields is set as $[-10,10]$. Then, the encoding map for QAEs can be obtained by $U_{\rm e} =U(T)$ following Eq.~(\ref{eq: unitary propagator}). 

\subsection{Training QAEs using learning algorithms}\label{subsec:train}

In this work, the training of QAEs is reduced to searching for an optimal $U_{\rm e} $ that maximizes $\Phi({\rm w})={\rm w} J_{\rm{pure}}+(1-{\rm w})J_{\rm{qmi}}, {\rm w}\in[0,1]$. After the training is completed, injecting a mixed state $\rho_{\rm{ref}}$ to the decoder helps maintain the entropy consistency between the initial state $\rho_0$ and the recovered state $\rho_f = U_{\rm e} ^{\dagger} (\rho_{\rm latent} \otimes \rho_{\rm ref})  U_{\rm e} $.  Finally, the overlap between the recovered states and the original states is measured to evaluate the efficiency of QAEs. Denote the parameters for the encoder as a vector $\boldsymbol{\theta}$. The procedure of QAE-qmi using mixed reference states is as follows:
\begin{enumerate}
	\item Randomly initialize $\boldsymbol{\theta}$, where $\boldsymbol{\theta}$ represents the control fields for the systems
	\item Apply $U_{\rm e} (\boldsymbol{\theta})$ to the initial states $\rho_0$
	\item Measure $\rho_{\rm{trash}}$ and $\rho_{\rm latent}$ and compute the cost function \\$\Phi({\rm w})={\rm w} F(\rho_{\rm{trash}},|0\rangle \langle 0| )-(1-{\rm w}) \mathcal{I}(U_{\rm e}  \rho_0 U_{\rm e} ^{\dagger})$
	\item Perform the optimization of $\Phi(\rm w)$ using a learning algorithm and obtain a better control parameter $\boldsymbol{\theta}$
	\item Repeat steps 2-4 until convergence
	\item Report the classical information $\boldsymbol{\theta}$ and store the latent state $\rho_{\rm latent}$
	\item Determine a suitable reference state $\rho_{\rm{ref}}$ using different strategies (e.g., $\rho_{\rm{ref}}=\rho_{\rm{trash}}$ or $\rho_{\rm{ref}}=\rho_{\rm mix}$) and prepare the reference state 
	\item Perform $U_{\rm e} ^{\dagger}(\boldsymbol{\theta})$ on the combined state $ \rho_{\rm latent} \otimes \rho_{\rm{ref}} $ and obtain the recovered state as $\rho_f$
\end{enumerate}

The key is to optimize the cost function of $\Phi$ using learning algorithms. Evolutionary strategy (ES) methods exhibit an advantage in exploring unknown environments in games~\cite{salimans2017evolution} and have been applied in optimizing quantum control issues~\cite{shir2009niching}. The comparison results in our previous work suggest that ES has the potential to optimize QAEs towards the theoretical upper bounds with high efficiency~\cite{ma2023compression}. In this work, we utilize ES to optimize the cost function $\Phi(\rm w)$.

ES is a black-box optimization method that utilizes heuristic search procedures inspired by natural evolution. At every iteration (``generation”), a population of parameter vectors (``genotypes”) is perturbed (``mutated”) and their objective function value (``fitness”) is evaluated. The highest-scoring parameter vectors are then recombined to form the population for the next generation, and this procedure is iterated until convergence~\cite{salimans2017evolution}. The detailed description for the ES method is provided in Supplementary Method 1.

It is worth noting that, the initialization process in Step 1 can be formulated as  $\boldsymbol{\theta} =\boldsymbol{u}_{\textup{min}} +\textup{Rand}[0,1](\boldsymbol{u}_{\textup{max}}-\boldsymbol{u}_{\textup{min}})$, where $\textup{Rand}[0,1]$ is a function to generate random numbers uniformly distributed between 0 and 1 to meet the physical restriction of control fields. In addition,  boundary checks and resetting values are required for every step that involves new parameters to guarantee that newly generated parameters lie in the constrained field.  For the parameter setting of the ES method, we set the population size as $\rm{NP}=40$ for 2-qubit states and $\rm{NP}=50$ for 4-qubit states. The perturbation factor is set as $\delta=0.01$. The learning rate is set as $\chi_1=0.5$. The momentum factor is set as $\chi_2=0.9$. The perturbation factor is decayed as $\delta \leftarrow 0.98\delta$ every 100 training iterations. 

\section{Data Availability Statement}
The data generated in this study have been deposited in the Figshare database, which can be accessed by 10.6084/m9.figshare.25183358.

\section{Acknowledgements}
This work was supported by the Australian Research
Council's Future Fellowship funding scheme under Project FT220100656, the Australian Research Council’s Discovery Projects funding scheme DP210101938, and the University of Melbourne through the establishment of the IBM Quantum Network Hub at the University.

H.M. and D.D. would like to thank Yuanlong Wang and Shuixin Xiao for helpful discussions. 

\section{Competing Interests}
The authors declare that they have no competing interests.

\section{Author Contributions}
H.M., D.D., and I.R.P. developed the theory. H.M. performed the numerical simulation and analyzed  data with the assistance of D.D. and I.R.P. H.M., G.J.M., and L.C.L.H. designed the experiment. H.M. and G.J.M. analyzed the experimental data with the help of other authors. All authors contributed in writing the paper.

\bibliographystyle{naturemag}  
\bibliography{bib/reference}

\begin{thebibliography}{10}
\expandafter\ifx\csname url\endcsname\relax
  \def\url#1{\texttt{#1}}\fi
\expandafter\ifx\csname urlprefix\endcsname\relax\def\urlprefix{URL }\fi
\providecommand{\bibinfo}[2]{#2}
\providecommand{\eprint}[2][]{\url{#2}}

\bibitem{biamonte2017quantum}
\bibinfo{author}{Biamonte, J.} \emph{et~al.}
\newblock \bibinfo{title}{Quantum machine learning}.
\newblock \emph{\bibinfo{journal}{Nature}} \textbf{\bibinfo{volume}{549}},
  \bibinfo{pages}{195} (\bibinfo{year}{2017}).

\bibitem{dong2008quantum}
\bibinfo{author}{Dong, D.}, \bibinfo{author}{Chen, C.}, \bibinfo{author}{Li,
  H.} \& \bibinfo{author}{Tarn, T.-J.}
\newblock \bibinfo{title}{Quantum reinforcement learning}.
\newblock \emph{\bibinfo{journal}{IEEE Trans. Syst. Man. Cybern, Part B
  (Cybernetics)}} \textbf{\bibinfo{volume}{38}}, \bibinfo{pages}{1207--1220}
  (\bibinfo{year}{2008}).

\bibitem{huang2021power}
\bibinfo{author}{Huang, H.-Y.} \emph{et~al.}
\newblock \bibinfo{title}{Power of data in quantum machine learning}.
\newblock \emph{\bibinfo{journal}{Nat. Commun.}} \textbf{\bibinfo{volume}{12}},
  \bibinfo{pages}{2631} (\bibinfo{year}{2021}).

\bibitem{cerezo2022challenges}
\bibinfo{author}{Cerezo, M.}, \bibinfo{author}{Verdon, G.},
  \bibinfo{author}{Huang, H.-Y.}, \bibinfo{author}{Cincio, L.} \&
  \bibinfo{author}{Coles, P.~J.}
\newblock \bibinfo{title}{Challenges and opportunities in quantum machine
  learning}.
\newblock \emph{\bibinfo{journal}{Nat. Comput. Sci.}}
  \textbf{\bibinfo{volume}{2}}, \bibinfo{pages}{567--576}
  (\bibinfo{year}{2022}).

\bibitem{niu2019universal}
\bibinfo{author}{Niu, M.~Y.}, \bibinfo{author}{Boixo, S.},
  \bibinfo{author}{Smelyanskiy, V.~N.} \& \bibinfo{author}{Neven, H.}
\newblock \bibinfo{title}{Universal quantum control through deep reinforcement
  learning}.
\newblock \emph{\bibinfo{journal}{npj Quantum Inf.}}
  \textbf{\bibinfo{volume}{5}}, \bibinfo{pages}{33} (\bibinfo{year}{2019}).

\bibitem{li2020quantum}
\bibinfo{author}{Li, J.-A.} \emph{et~al.}
\newblock \bibinfo{title}{Quantum reinforcement learning during human
  decision-making}.
\newblock \emph{\bibinfo{journal}{Nat. Hum. Behav.}}
  \textbf{\bibinfo{volume}{4}}, \bibinfo{pages}{294--307}
  (\bibinfo{year}{2020}).

\bibitem{dong2022quantum}
\bibinfo{author}{Dong, D.} \& \bibinfo{author}{Petersen, I.~R.}
\newblock \bibinfo{title}{Quantum estimation, control and learning:
  opportunities and challenges}.
\newblock \emph{\bibinfo{journal}{Annu. Rev. Control.}}
  \textbf{\bibinfo{volume}{54}}, \bibinfo{pages}{243--251}
  (\bibinfo{year}{2022}).

\bibitem{pu2016variational}
\bibinfo{author}{Pu, Y.} \emph{et~al.}
\newblock \bibinfo{title}{Variational autoencoder for deep learning of images,
  labels and captions}.
\newblock In \emph{\bibinfo{booktitle}{Adv. Neural Inform. Process. Syst.}},
  \bibinfo{pages}{2352--2360} (\bibinfo{year}{2016}).

\bibitem{bartuuvskova2006optical}
\bibinfo{author}{Bart\r{u}\v{s}kov\'{a}, L.} \emph{et~al.}
\newblock \bibinfo{title}{Optical implementation of the encoding of two qubits
  to a single qutrit}.
\newblock \emph{\bibinfo{journal}{Phys. Rev. A}} \textbf{\bibinfo{volume}{74}},
  \bibinfo{pages}{022325} (\bibinfo{year}{2006}).

\bibitem{steinbrecher2019quantum}
\bibinfo{author}{Steinbrecher, G.~R.}, \bibinfo{author}{Olson, J.~P.},
  \bibinfo{author}{Englund, D.} \& \bibinfo{author}{Carolan, J.}
\newblock \bibinfo{title}{Quantum optical neural networks}.
\newblock \emph{\bibinfo{journal}{npj Quantum Inf.}}
  \textbf{\bibinfo{volume}{5}}, \bibinfo{pages}{60} (\bibinfo{year}{2019}).

\bibitem{lamata2018quantum}
\bibinfo{author}{Lamata, L.}, \bibinfo{author}{Alvarez-Rodriguez, U.},
  \bibinfo{author}{Mart{\'\i}n-Guerrero, J.~D.}, \bibinfo{author}{Sanz, M.} \&
  \bibinfo{author}{Solano, E.}
\newblock \bibinfo{title}{Quantum autoencoders via quantum adders with genetic
  algorithms}.
\newblock \emph{\bibinfo{journal}{Quantum Mach. Learn.: Sci. Technol.}}
  \textbf{\bibinfo{volume}{4}}, \bibinfo{pages}{014007} (\bibinfo{year}{2018}).

\bibitem{ding2019experimental}
\bibinfo{author}{Ding, Y.}, \bibinfo{author}{Lamata, L.},
  \bibinfo{author}{Sanz, M.}, \bibinfo{author}{Chen, X.} \&
  \bibinfo{author}{Solano, E.}
\newblock \bibinfo{title}{Experimental implementation of a quantum autoencoder
  via quantum adders}.
\newblock \emph{\bibinfo{journal}{Adv. Quantum Technol.}}
  \textbf{\bibinfo{volume}{2}}, \bibinfo{pages}{1800065}
  (\bibinfo{year}{2019}).

\bibitem{aspuru2005simulated}
\bibinfo{author}{Aspuru-Guzik, A.}, \bibinfo{author}{Dutoi, A.~D.},
  \bibinfo{author}{Love, P.~J.} \& \bibinfo{author}{Head-Gordon, M.}
\newblock \bibinfo{title}{Simulated quantum computation of molecular energies}.
\newblock \emph{\bibinfo{journal}{Science}} \textbf{\bibinfo{volume}{309}},
  \bibinfo{pages}{1704--1707} (\bibinfo{year}{2005}).

\bibitem{wan2017quantum}
\bibinfo{author}{Wan, K.~H.}, \bibinfo{author}{Dahlsten, O.},
  \bibinfo{author}{Kristj{\'a}nsson, H.}, \bibinfo{author}{Gardner, R.} \&
  \bibinfo{author}{Kim, M.}
\newblock \bibinfo{title}{Quantum generalisation of feedforward neural
  networks}.
\newblock \emph{\bibinfo{journal}{npj Quantum Inf.}}
  \textbf{\bibinfo{volume}{3}}, \bibinfo{pages}{36} (\bibinfo{year}{2017}).

\bibitem{romero2017quantum}
\bibinfo{author}{Romero, J.}, \bibinfo{author}{Olson, J.~P.} \&
  \bibinfo{author}{Aspuru-Guzik, A.}
\newblock \bibinfo{title}{Quantum autoencoders for efficient compression of
  quantum data}.
\newblock \emph{\bibinfo{journal}{Quantum Sci. Technol.}}
  \textbf{\bibinfo{volume}{2}}, \bibinfo{pages}{045001} (\bibinfo{year}{2017}).

\bibitem{bravo2021quantum}
\bibinfo{author}{Bravo-Prieto, C.}
\newblock \bibinfo{title}{Quantum autoencoders with enhanced data encoding}.
\newblock \emph{\bibinfo{journal}{Mach. Learn.: Sci. Technol.}}
  \textbf{\bibinfo{volume}{2}}, \bibinfo{pages}{035028} (\bibinfo{year}{2021}).

\bibitem{pepper2019experimental}
\bibinfo{author}{Pepper, A.}, \bibinfo{author}{Tischler, N.} \&
  \bibinfo{author}{Pryde, G.~J.}
\newblock \bibinfo{title}{Experimental realization of a quantum autoencoder:
  The compression of qutrits via machine learning}.
\newblock \emph{\bibinfo{journal}{Phys. Rev. Lett.}}
  \textbf{\bibinfo{volume}{122}}, \bibinfo{pages}{060501}
  (\bibinfo{year}{2019}).

\bibitem{huang2020realization}
\bibinfo{author}{Huang, C.-J.} \emph{et~al.}
\newblock \bibinfo{title}{Realization of a quantum autoencoder for lossless
  compression of quantum data}.
\newblock \emph{\bibinfo{journal}{Phys. Rev. A}}
  \textbf{\bibinfo{volume}{102}}, \bibinfo{pages}{032412}
  (\bibinfo{year}{2020}).

\bibitem{bondarenko2020quantum}
\bibinfo{author}{Bondarenko, D.} \& \bibinfo{author}{Feldmann, P.}
\newblock \bibinfo{title}{Quantum autoencoders to denoise quantum data}.
\newblock \emph{\bibinfo{journal}{Phys. Rev. Lett.}}
  \textbf{\bibinfo{volume}{124}}, \bibinfo{pages}{130502}
  (\bibinfo{year}{2020}).

\bibitem{achache2020denoising}
\bibinfo{author}{Achache, T.}, \bibinfo{author}{Horesh, L.} \&
  \bibinfo{author}{Smolin, J.}
\newblock \bibinfo{title}{Denoising quantum states with quantum
  autoencoders--theory and applications}.
\newblock \emph{\bibinfo{journal}{Preprint at
  \textup{[https://arxiv.org/pdf/2012.14714]}}}  (\bibinfo{year}{2020}).

\bibitem{zhang2021generic}
\bibinfo{author}{Zhang, X.-M.} \emph{et~al.}
\newblock \bibinfo{title}{Generic detection-based error mitigation using
  quantum autoencoders}.
\newblock \emph{\bibinfo{journal}{Phys. Rev. A}}
  \textbf{\bibinfo{volume}{103}}, \bibinfo{pages}{L040403}
  (\bibinfo{year}{2021}).

\bibitem{du2021exploring}
\bibinfo{author}{Du, Y.} \& \bibinfo{author}{Tao, D.}
\newblock \bibinfo{title}{On exploring practical potentials of quantum
  auto-encoder with advantages}.
\newblock \emph{\bibinfo{journal}{Preprint at
  \textup{[https://arxiv.org/pdf/2106.15432]}}}  (\bibinfo{year}{2021}).

\bibitem{srikumar2021clustering}
\bibinfo{author}{Srikumar, M.}, \bibinfo{author}{Hill, C.~D.} \&
  \bibinfo{author}{Hollenberg, L.~C.}
\newblock \bibinfo{title}{Clustering and enhanced classification using a hybrid
  quantum autoencoder}.
\newblock \emph{\bibinfo{journal}{Quantum Mach. Learn.: Sci. Technol.}}
  \textbf{\bibinfo{volume}{7}}, \bibinfo{pages}{015020} (\bibinfo{year}{2021}).

\bibitem{zhang2022resource}
\bibinfo{author}{Zhang, H.} \emph{et~al.}
\newblock \bibinfo{title}{Resource-efficient high-dimensional subspace
  teleportation with a quantum autoencoder}.
\newblock \emph{\bibinfo{journal}{Sci. Adv.}} \textbf{\bibinfo{volume}{8}},
  \bibinfo{pages}{9783} (\bibinfo{year}{2022}).

\bibitem{mangini2022quantum}
\bibinfo{author}{Mangini, S.} \emph{et~al.}
\newblock \bibinfo{title}{Quantum neural network autoencoder and classifier
  applied to an industrial case study}.
\newblock \emph{\bibinfo{journal}{Quantum Mach. Intell.}}
  \textbf{\bibinfo{volume}{4}}, \bibinfo{pages}{13} (\bibinfo{year}{2022}).

\bibitem{ma2023compression}
\bibinfo{author}{Ma, H.} \emph{et~al.}
\newblock \bibinfo{title}{On compression rate of quantum autoencoders: Control
  design, numerical and experimental realization}.
\newblock \emph{\bibinfo{journal}{Automatica}} \textbf{\bibinfo{volume}{147}},
  \bibinfo{pages}{110659} (\bibinfo{year}{2023}).

\bibitem{cao2021noise}
\bibinfo{author}{Cao, C.} \& \bibinfo{author}{Wang, X.}
\newblock \bibinfo{title}{Noise-assisted quantum autoencoder}.
\newblock \emph{\bibinfo{journal}{Phys. Rev. Appl.}}
  \textbf{\bibinfo{volume}{15}}, \bibinfo{pages}{054012}
  (\bibinfo{year}{2021}).

\bibitem{pivoluska2022implementation}
\bibinfo{author}{Pivoluska, M.} \& \bibinfo{author}{Plesch, M.}
\newblock \bibinfo{title}{Implementation of quantum compression on {IBM}
  quantum computers}.
\newblock \emph{\bibinfo{journal}{Sci. Rep.}} \textbf{\bibinfo{volume}{12}},
  \bibinfo{pages}{5841} (\bibinfo{year}{2022}).

\bibitem{nielsen2010quantum}
\bibinfo{author}{Nielsen, M.~A.} \& \bibinfo{author}{Chuang, I.~L.}
\newblock \emph{\bibinfo{title}{Quantum Computation and Quantum Information}}
  (\bibinfo{publisher}{Cambridge University Press}, \bibinfo{year}{2010}).

\bibitem{wilde2011classical}
\bibinfo{author}{Wilde, M.~M.}
\newblock \bibinfo{title}{From classical to quantum {Shannon} theory}.
\newblock \emph{\bibinfo{journal}{Preprint at
  \textup{[https://arxiv.org/pdf/1106.1445]}}}  (\bibinfo{year}{2011}).

\bibitem{watrous2018theory}
\bibinfo{author}{Watrous, J.}
\newblock \emph{\bibinfo{title}{The Theory of Quantum Information}}
  (\bibinfo{publisher}{Cambridge University Press}, \bibinfo{year}{2018}).

\bibitem{lyons2012werner}
\bibinfo{author}{Lyons, D.~W.}, \bibinfo{author}{Skelton, A.~M.} \&
  \bibinfo{author}{Walck, S.~N.}
\newblock \bibinfo{title}{Werner state structure and entanglement
  classification}.
\newblock \emph{\bibinfo{journal}{Adv. Math. Phys.}}
  \textbf{\bibinfo{volume}{2012}} (\bibinfo{year}{2012}).

\bibitem{ibmQST2021}
\bibinfo{title}{Qiskit library for quantum state tomography,
  https://qiskit.org/ecosystem/experiments/stubs/\\qiskit$\_$experiments.library.tomography.\\statetomography.html}
  (\bibinfo{year}{2021}).

\bibitem{dong2010quantum}
\bibinfo{author}{Dong, D.} \& \bibinfo{author}{Petersen, I.~R.}
\newblock \bibinfo{title}{Quantum control theory and applications: a survey}.
\newblock \emph{\bibinfo{journal}{IET Control Theory Appl.}}
  \textbf{\bibinfo{volume}{4}}, \bibinfo{pages}{2651--2671}
  (\bibinfo{year}{2010}).

\bibitem{burgarth2009local}
\bibinfo{author}{Burgarth, D.}, \bibinfo{author}{Bose, S.},
  \bibinfo{author}{Bruder, C.} \& \bibinfo{author}{Giovannetti, V.}
\newblock \bibinfo{title}{Local controllability of quantum networks}.
\newblock \emph{\bibinfo{journal}{Phys. Rev. A}} \textbf{\bibinfo{volume}{79}},
  \bibinfo{pages}{060305} (\bibinfo{year}{2009}).

\bibitem{wang2016subspace}
\bibinfo{author}{Wang, X.}, \bibinfo{author}{Burgarth, D.} \&
  \bibinfo{author}{Schirmer, S.}
\newblock \bibinfo{title}{Subspace controllability of spin-$1/2$ chains with
  symmetries}.
\newblock \emph{\bibinfo{journal}{Phys. Rev. A}} \textbf{\bibinfo{volume}{94}},
  \bibinfo{pages}{052319} (\bibinfo{year}{2016}).

\bibitem{dong2023learning}
\bibinfo{author}{Dong, D.} \& \bibinfo{author}{Petersen, I.~R.}
\newblock \emph{\bibinfo{title}{Learning and Robust Control in Quantum
  Technology}} (\bibinfo{publisher}{Springer Nature, Switzerland AG},
  \bibinfo{year}{2023}).

\bibitem{salimans2017evolution}
\bibinfo{author}{Salimans, T.}, \bibinfo{author}{Ho, J.},
  \bibinfo{author}{Chen, X.}, \bibinfo{author}{Sidor, S.} \&
  \bibinfo{author}{Sutskever, I.}
\newblock \bibinfo{title}{Evolution strategies as a scalable alternative to
  reinforcement learning}.
\newblock \emph{\bibinfo{journal}{Preprint at
  \textup{[https://arxiv.org/pdf/1703.03864]}}}  (\bibinfo{year}{2017}).

\bibitem{shir2009niching}
\bibinfo{author}{Shir, O.~M.} \& \bibinfo{author}{B{\"a}ck, T.}
\newblock \bibinfo{title}{Niching with derandomized evolution strategies in
  artificial and real-world landscapes}.
\newblock \emph{\bibinfo{journal}{Nat. Comput.}} \textbf{\bibinfo{volume}{8}},
  \bibinfo{pages}{171--196} (\bibinfo{year}{2009}).

\bibitem{shende2005synthesis}
\bibinfo{author}{Shende, V.~V.}, \bibinfo{author}{Bullock, S.~S.} \&
  \bibinfo{author}{Markov, I.~L.}
\newblock \bibinfo{title}{Synthesis of quantum logic circuits}.
\newblock In \emph{\bibinfo{booktitle}{Proc. ASP-DAC 2005}},
  \bibinfo{pages}{272--275} (\bibinfo{year}{2005}).

\end{thebibliography}

\appendix 

	\section{Supplementary Method 1. Description of ES}\label{sup:es} 
Denote the control fields as a column vector $\theta$ and denote $\Phi(\theta)$ the loss function. Given the population size $NP$, the learning rate $\chi_1$, the momentum coefficient $\chi_2$, and the permutation factor $\delta$, the procedure of ES is as follows.
\begin{enumerate}
	\item Randomly initialize the mean vector $\bar{\theta}$
	\item Initialize the gradient $d\Phi=0$ and momentum $dv=0$
	\item Repeat for each individual $i=1,...,NP$
	\begin{enumerate}
		\item Sample variation $\epsilon_{i} \sim N(0,I)$
		\item Set mutation variant as $X_i \leftarrow \bar{\theta}+ \delta \epsilon_i $
	\end{enumerate}
	\item Obtain the gradient $ d\Phi \leftarrow \frac{1}{NP\delta}\sum_{i=1}^{NP}\Phi(X_i)\epsilon_i$
	\item Obtain the momentum $dv \leftarrow \chi_2 dv +(1-\chi_2) d\Phi$
	\item Update the new mean vector $\bar{\theta} \leftarrow \bar{\theta} + \chi_1 dv $
	\item If convergent, go to 8; otherwise, go to 3
	\item Optimal control parameters $\theta^*=\bar{\theta}$
\end{enumerate}

\section{Supplementary Discussion 1: Density matrices of 2-qubit states}

\begin{figure*}[htb]
	\centering
	
	\includegraphics[width=7.0in]{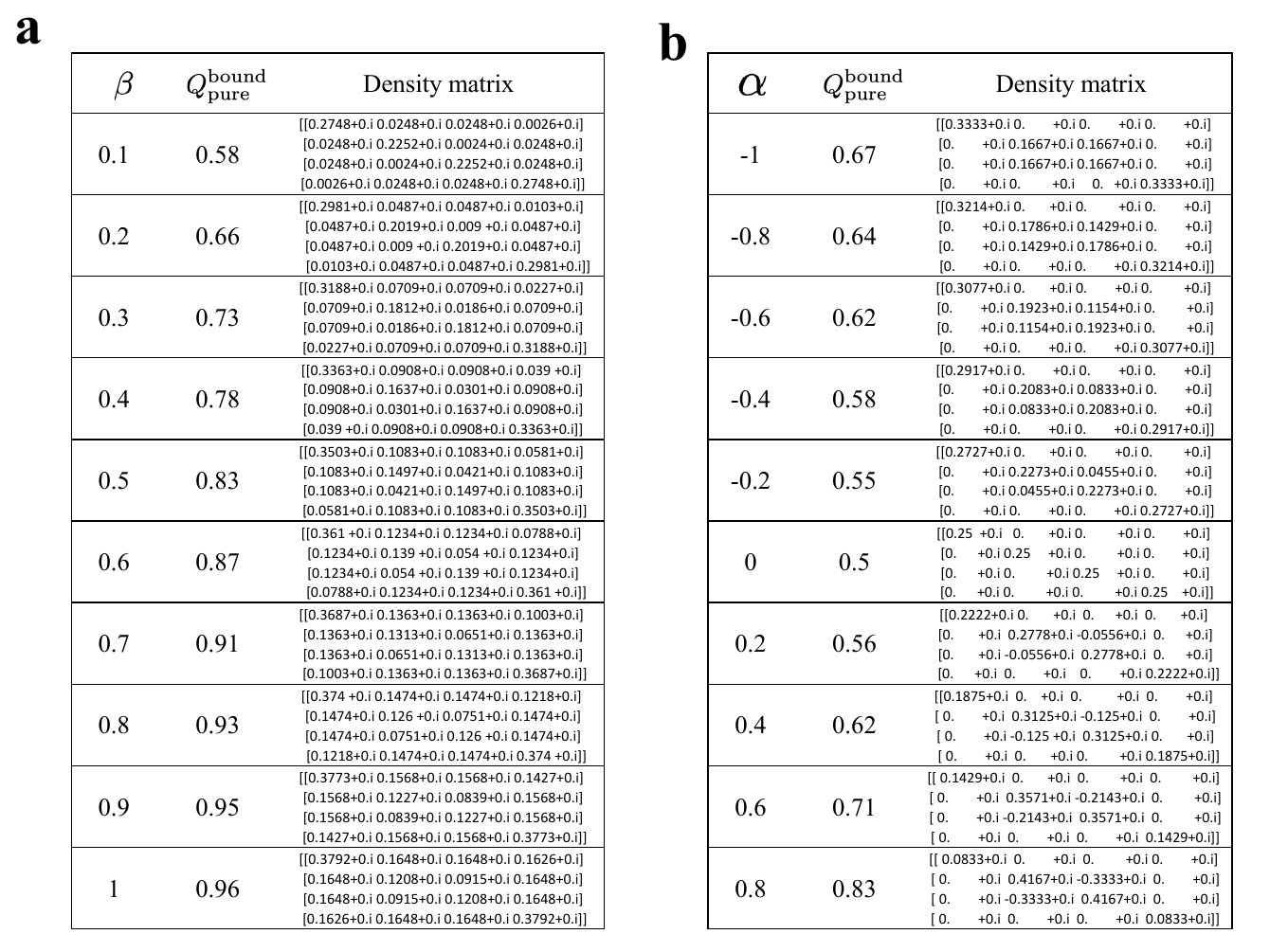}
	\caption{Density matrices for 2-qubit states (to 4 decimal places). \textbf{a} for thermal states, \textbf{b} for Werner states.}
	\label{fig:2-qubitdensity}
\end{figure*}

The density matrices for thermal states and Werner states are given in Fig.~\ref{fig:2-qubitdensity}. 

For quantum states in the form of Eq.~(8), we choose a random pure state and utilize the same $|\psi\rangle$ for different $p_0$. For the 2-qubit case, we take the following vector {\scriptsize
	\[
	[ 0.7085 - 0.2668j,\ 0.1391 - 0.1087j,\ -0.0752 - 0.3829j,\ 0.3929 - 0.2982j]^T
	\]
}
For the 4-qubit case, we choose the following vector
\[
\resizebox{\linewidth}{!}{$
	\begin{aligned}
		[0.2804 - 0.1056j,\ -0.1186 - 0.1459j,\ 0.0843 - 0.0130j,\ -0.0593 + 0.0075j, \\
		-0.2070 + 0.1490j,\ -0.0892 - 0.1721j,\ 0.3288 + 0.1918j,\ -0.2105 + 0.4228j,\\
		0.3541 + 0.1011j,\ -0.1927 + 0.0139j,\ -0.0234 - 0.1376j,\ -0.0336 + 0.1390j,\\
		0.1066 - 0.0317j,\ -0.1342 + 0.1926j,\ -0.0664 + 0.2517j,\ 0.0889 - 0.1924j]^T.
	\end{aligned}
	$}
\]

\section{Supplementary Discussion 2: Training performance of $\rho_{\rm ref}=\rho_{\rm trash}$}\label{sup:train}

\begin{figure*}[htb]
	\centering
	\includegraphics[width=7.0in]{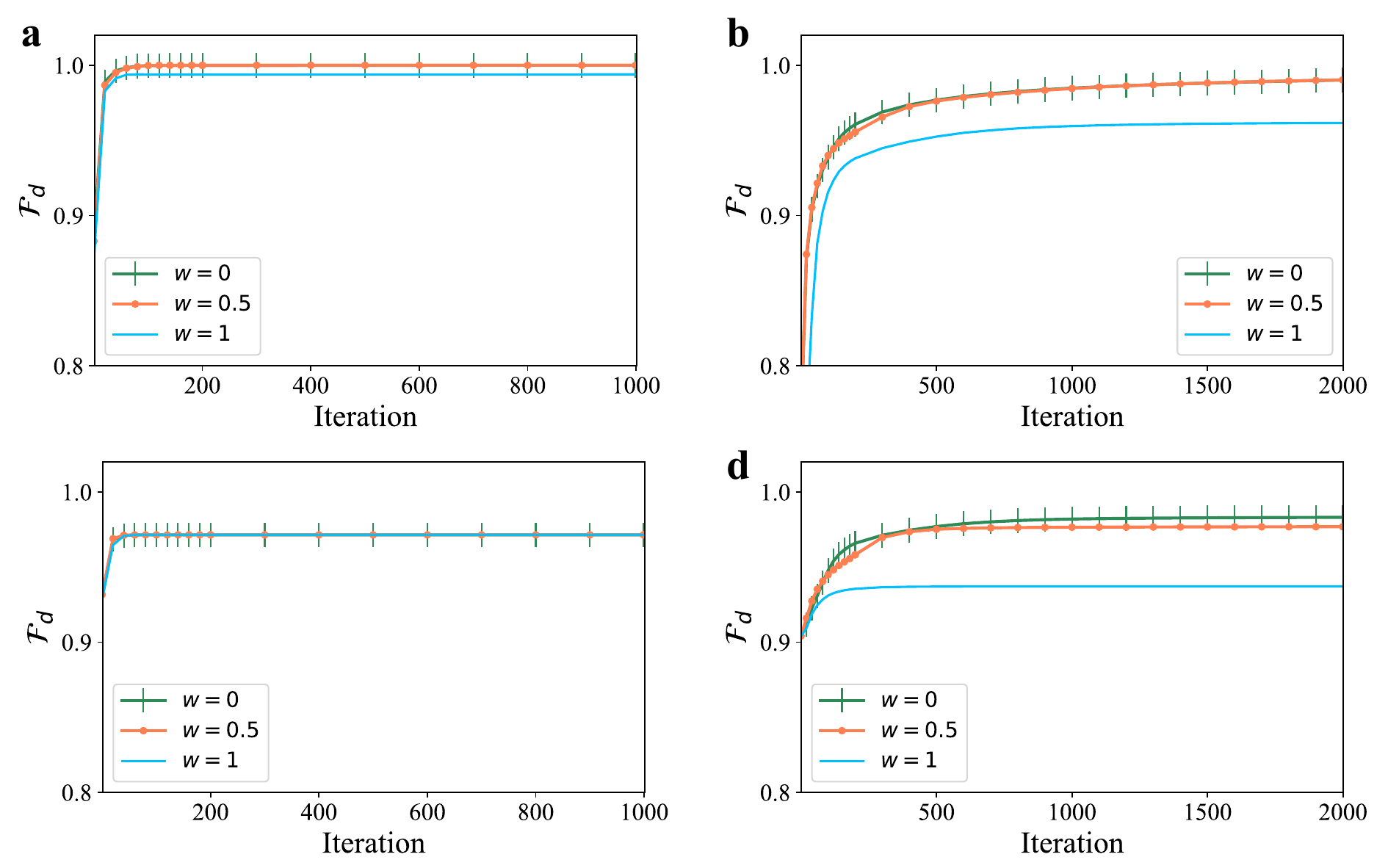}
	\caption{Training process of QAE-qmi using $\rho_{\rm ref}=\rho_{\rm trash}$ under different $\rm w$ for thermal states and Werner states. \textbf{a} for 2-qubit thermal states,  \textbf{b} for 4-qubit thermal states,  \textbf{c} for 2-qubit Werner states,  \textbf{d} for 4-qubit Werner states. $(1-\rm w)$ denotes the ratio of QMI in the new cost function $\Phi(\rm w)$. $\mathcal{F}_d$ denotes the decoding fidelity between the initial and recovered states.}
	\label{fig:Jab-train}
\end{figure*}

\begin{figure*}[htb]
	\centering
	\includegraphics[width=7.0in]{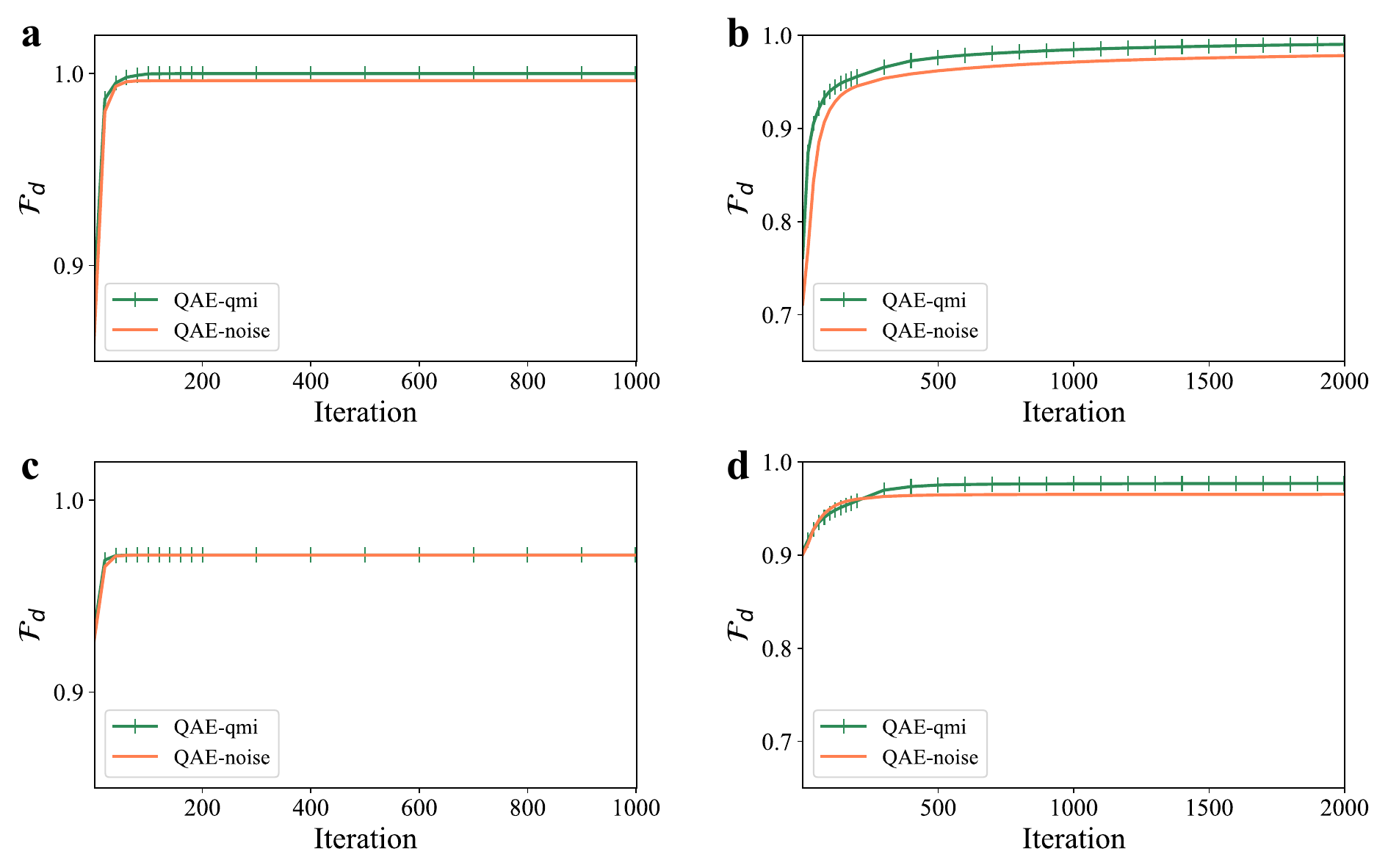}
	\caption{Training process of QAE-qmi vs QAE-noise. \textbf{a} for 2-qubit thermal states,  \textbf{b} for 4-qubit thermal states,  \textbf{c} for 2-qubit Werner states,  \textbf{d} for 4-qubit Werner states. QAE-qmi denotes the proposed method that takes $\rho_{\rm ref}=\rho_{\rm trash}$ and $\rm w=0$. QAE-noise denotes the method in \cite{cao2021noise}. $\mathcal{F}_d$ denotes the decoding fidelity between the initial state and the recovered state.} 
	\label{fig:ref=trash_train}
\end{figure*}

\begin{figure}[htb]
	\centering
	\includegraphics[width=7.0in]{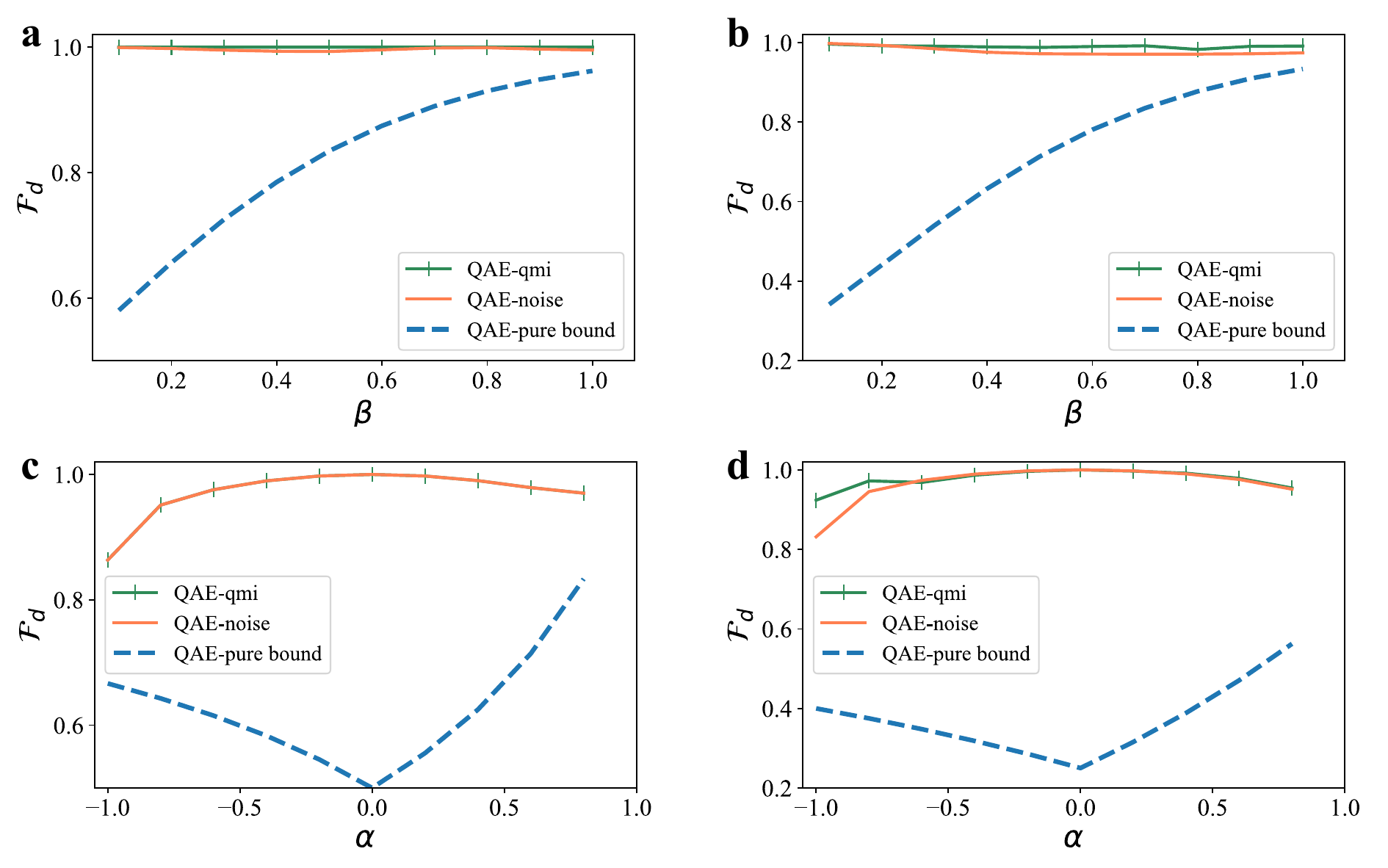}
	\caption{Comparison of QAE-qmi and QAE-noise for compressing different states. \textbf{a} for 2-qubit thermal states,  \textbf{b} for 4-qubit thermal states,  \textbf{c} for 2-qubit Werner states,  \textbf{d} for 4-qubit Werner states. $\mathcal{F}_d$ denotes the decoding fidelity between the initial and recovered states. The blue dashed line represents the theoretical upper bound of compression rate when only considering pure reference states~\cite{ma2023compression}.}
	\label{fig:qmi-noise_compare}
\end{figure}

First, we investigate the performance of different $\rm w $ in the cost function. The results are summarized in Fig. \ref{fig:Jab-train}, where $\rm w =0$ and $\rm w =0.5$ achieve a similar performance for both 2-qubit states and 4-qubit states. The gaps between $\rm w =0$ and $\rm w =1$ become large for 4-qubit states.  The way of $\rho_{\rm ref}=\rho_{\rm trash}$ is in line with maximizing $J_{\rm qmi}$. Hence, we set $\rm w =0$ with $\Phi=J_{\rm qmi}$ to investigate the performance of QAE using mixed reference states and compare it with the noise-assisted QAE method in \cite{cao2021noise} (termed as QAE-noise in this paper). 
The average training curves of the two methods are demonstrated in Fig. \ref{fig:ref=trash_train}, where they achieve similar decoding fidelities for 2-qubit states in (a1) and (b1). The gaps in (a2) and (b2) reveal that the advantage of QAE-qmi over QAE-noise becomes apparent for 4-qubit states.  The decoding fidelities for different parameters are summarized in Fig. \ref{fig:qmi-noise_compare}. The QAE-pure bound of thermal states increases with $\beta$, and QAE-qmi and QAE-noise both beat the QAE-pure bounds under different values of $\beta$.

\begin{figure}[htb]
	\centering
	\includegraphics[width=7.0in]{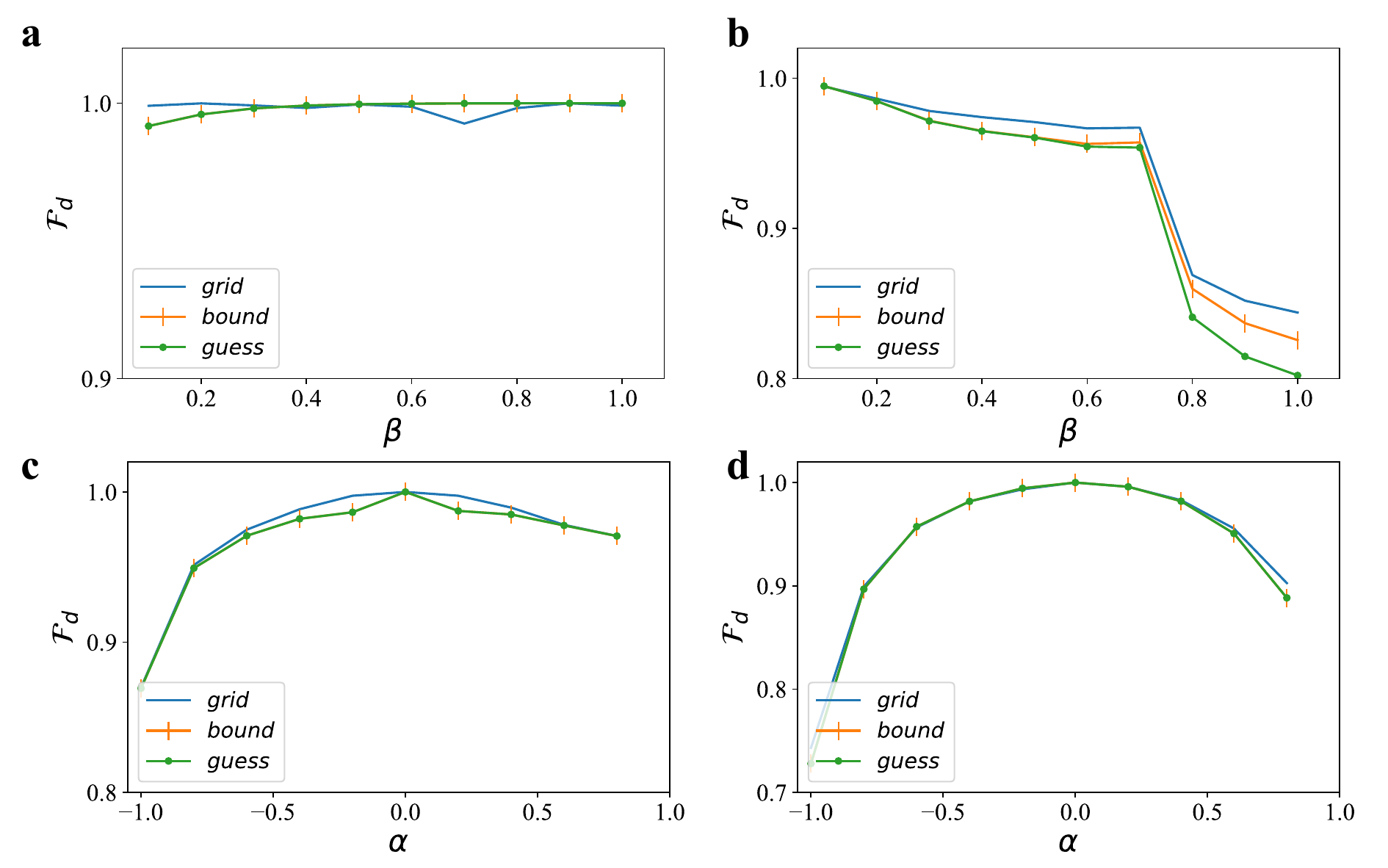}
	\caption{Comparison of different strategies for setting $p_r$ when compressing thermal and Werner states. \textbf{a} for 2-qubit thermal states,  \textbf{b} for 4-qubit thermal states,  \textbf{c} for 2-qubit Werner states,  \textbf{d} for 4-qubit Werner states. In the legends, grid denotes the strategy of setting $p_r$ using grid-search, bound denotes the strategy of setting $p_r$ by drawing inspirations from the QAE-pure bound, and guess denotes the way of setting $p_r$ by guessing (inferring) from the training process of QAE. $\mathcal{F}_d$ denotes the decoding fidelity between the initial state and the recovered state. }
	\label{fig:candip_result}
\end{figure}

\section{Supplementary Discussion 3: Different strategies of  $p_r$ for $\rho_{\rm ref}=\rho_{\rm mix}$}\label{sup:additional}

Numerical results of the three strategies for setting $p_r$ for compressing thermal states and Werner states are summarized in Fig. \ref{fig:candip_result}. As shown, the curves of $p_r^{\rm bound}$ and $p_r^{\rm guess}$ exhibit a similar performance, which is also comparable to that of $p_r^{\rm grid}$. Those results demonstrate that the manual and automatic strategies of setting $p_r$ are effective in compressing and recovering different quantum states.

\section{Supplementary Discussion 4: Different $\rm w $ under $\rho_{\rm ref}=\rho_{\rm mix}$}\label{sup:ratio}

\begin{figure*}[th]
	\centering
	\includegraphics[width=7.0in]{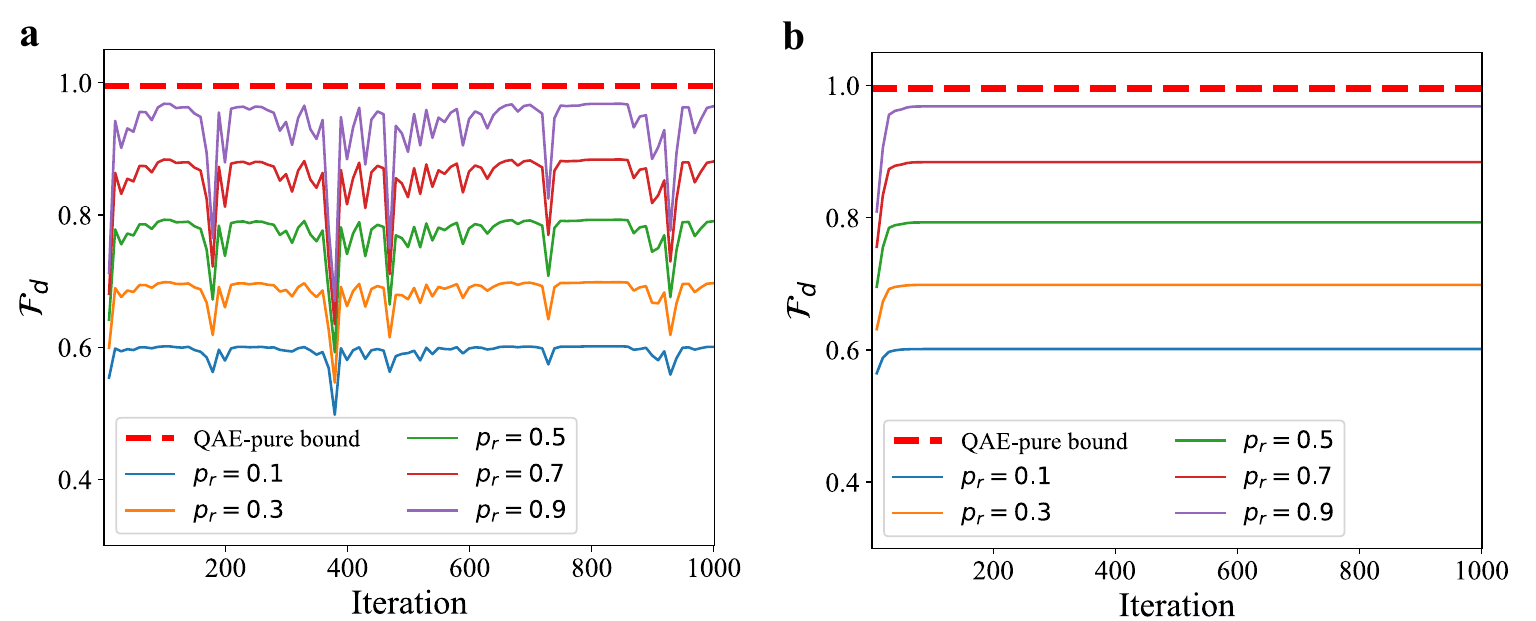}
	\caption{An example of QAE-qmi with limited reference states for compressing 2-qubit states with $p_0=0.99$. \textbf{a} for $\rm{w}=0.5$, \textbf{b} for $\rm w =0.99$. $p_r$ represents the ratio of the pure state and $(1-p_r)$ represents the ratio of the mixed state in the reference state. $\mathcal{F}_d$ represents the fidelity between the initial state and the recovered state. The red dashed line represents the theoretical upper bound of compression rate when only considering pure reference states~\cite{ma2023compression}. (a) training using $J=0.5(J_{\rm{pure}}+J_{\rm{qmi}})$, (b) training using $J=0.99J_{\rm{pure}}+0.01J_{\rm{qmi}}$.In this figure, some selected values of $p_r$ from the candidate pool are used to clearly demonstrate the trends of different values.}
	\label{fig:highpure-2qubit}
\end{figure*}

\begin{figure*}[h]
	\centering
	\includegraphics[width=7.0in]{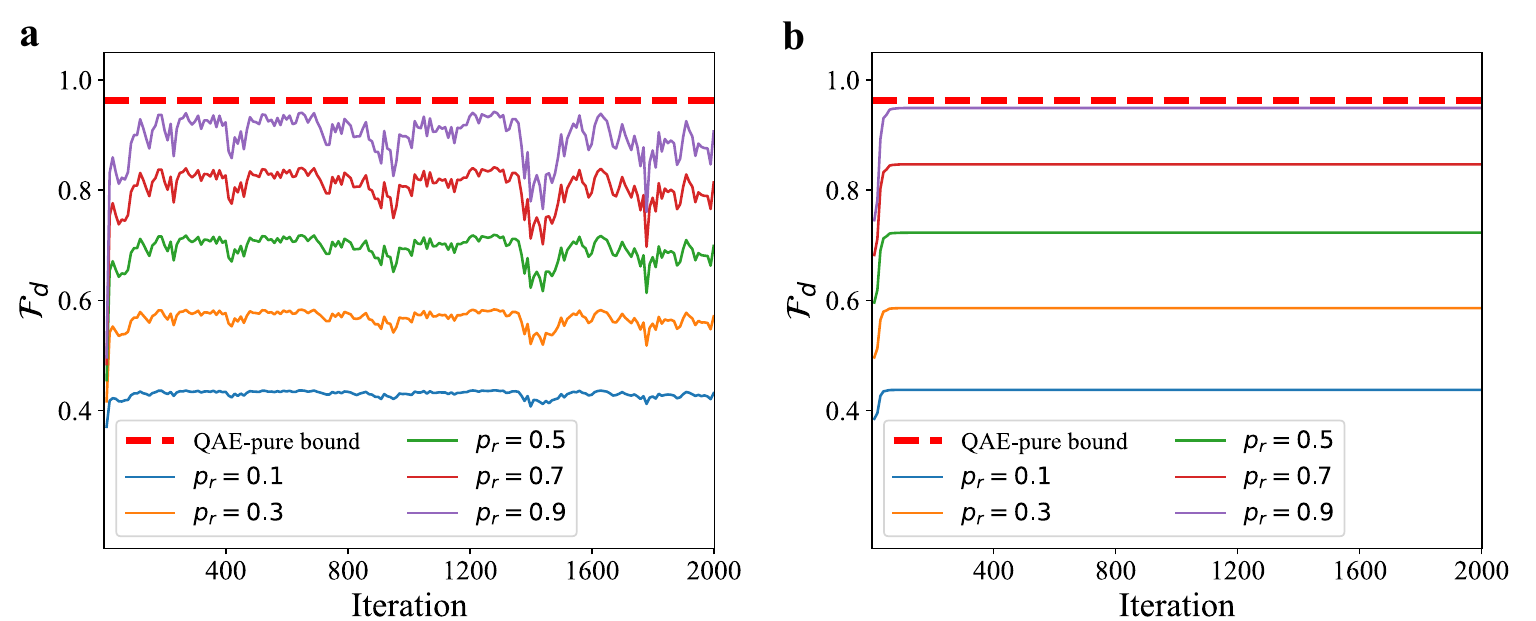}
	\caption{An example of QAE-qmi with limited reference states for compressing 4-qubit states with $p_0=0.95$. \textbf{a} for $\rm{w}=0.5$, \textbf{b} for $\rm w =0.99$. $p_r$ represents the ratio of the pure state and $(1-p_r)$ represents the ratio of the mixed state in the reference state. $\mathcal{F}_d$ represents the fidelity between the initial state and the recovered state. The red dashed line represents the theoretical upper bound of compression rate when only considering pure reference states~\cite{ma2023compression}. (a) training using $J=0.5(J_{\rm{pure}}+J_{\rm{qmi}})$, (b) training using $J=0.99J_{\rm{pure}}+0.01J_{\rm{qmi}}$. In this figure, some selected values of $p_r$ from the candidate pool are used to clearly demonstrate the trends of different values.}
	\label{fig:highpure-4qubit}
\end{figure*}

To better display this, we provide the training performance of QAE-qmi with $\rm w=0.5$ and $\rm w=0.99$ in  Fig.~\ref{fig:highpure-2qubit} for 2-qubit states and Fig.~\ref{fig:highpure-4qubit} for 4-qubit states. For states with high purities, training with $\Phi=0.5(J_{\rm{pure}}+J_{\rm{qmi}})$ might lead to fluctuation of decoding fidelity, which can be avoided by reducing the ratio of $J_{\rm{qmi}}$.

\begin{figure}[h]
	\centering
	\includegraphics[width=7.0in]{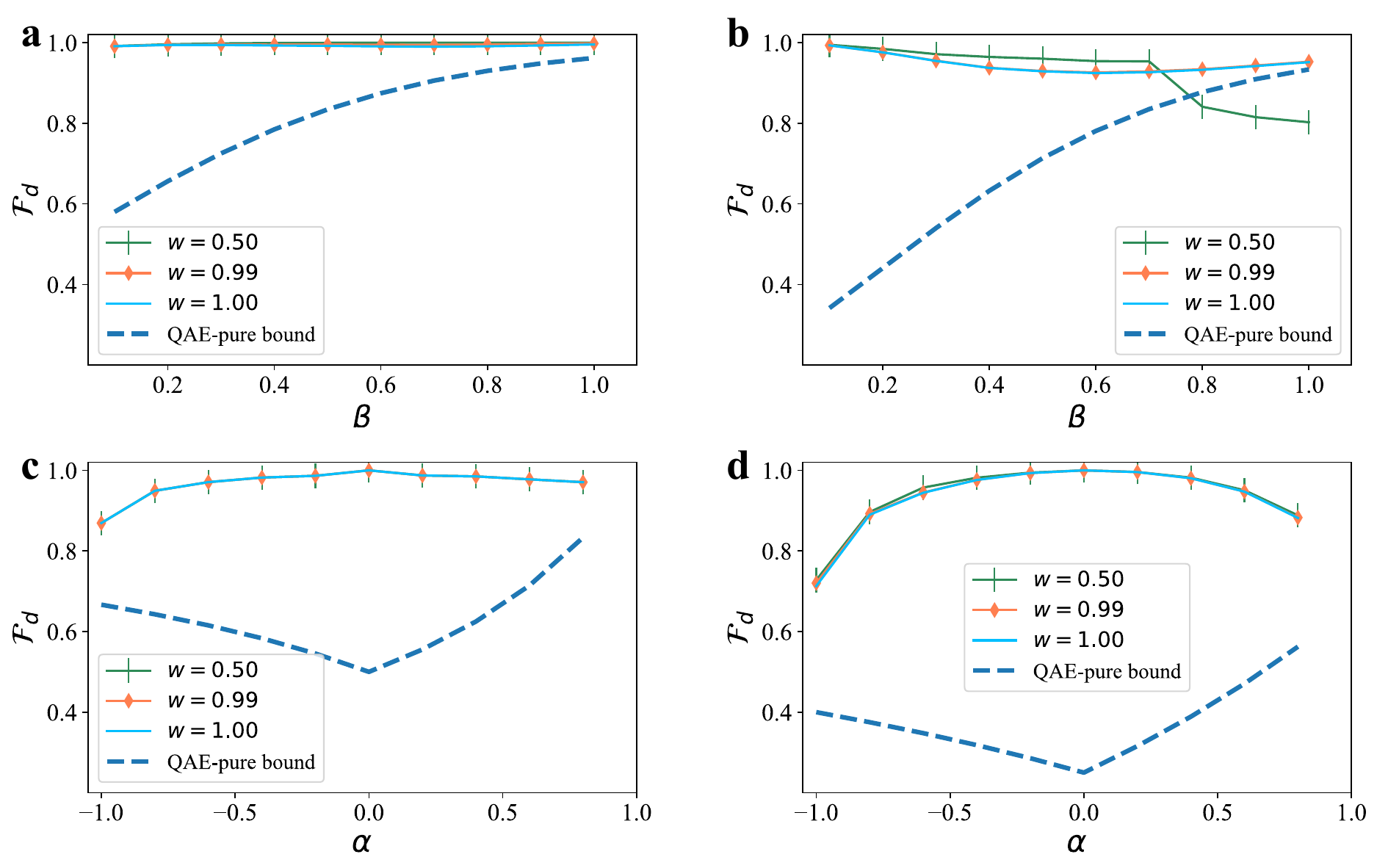}
	\caption{Comparison of different ratios in $\Phi(\rm w )$ when compressing thermal and Werner states under $p_r=p_r^{\rm guess}$. \textbf{a} for 2-qubit thermal states,  \textbf{b} for 4-qubit thermal states,  \textbf{c} for 2-qubit Werner states,  \textbf{d} for 4-qubit Werner states. $(1-\rm w)$ denotes the ratio of QMI in the new cost function $\Phi(\rm w)$. The blue dashed line represents the theoretical upper bound of compression rate when only considering pure reference states~\cite{ma2023compression}.}
	\label{fig:ratio_guess}
\end{figure}

\begin{figure}[h]
	\centering
	\includegraphics[width=7.0in]{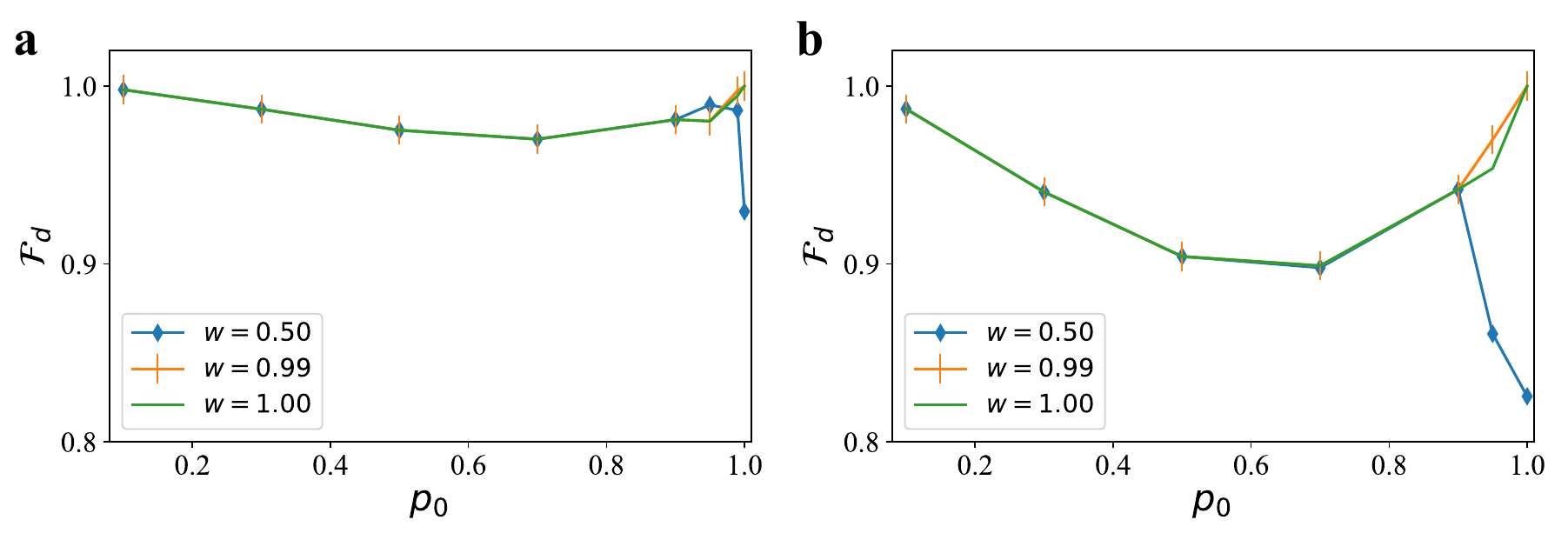}
	\caption{Decoding fidelity comparisons between different values of $\rm w $ for compressing maximally mixed states blended with pure states under $p_r=p_r^{\rm guess}$. \textbf{a} for 2-qubit states, \textbf{b} for 4-qubit states. $\rm w)$ denotes the ratio of $J_{\rm pure}$ in the cost function $\Phi(\rm w)$, and $\mathcal{F}_d$ represents the decoding fidelity between the initial state and the recovered state.}
	\label{fig:ratio_haar}
\end{figure}

In this section, we investigate the performance of $\rm w $ under the reference setting scheme $\rho_{\rm ref}=\rho_{\rm mix}$. We find that when compressing states with high QAE-pure bounds, it is necessary to introduce $J_{\rm pure}$ to allow adaptive mixedness to achieve good decoding fidelities. From Fig.~\ref{fig:candip_result}, under the reference scheme $\rho_{\rm ref}=\rho_{\rm mix}$, using $\Phi=0.5(J_{\rm pure}+J_{\rm qmi})$ fails to beat QAE-pure bounds for  4-qubit thermal states with large $\beta$ for all the three strategies of setting $p_r$. The reason is that 4-qubit thermal states with high $\beta$ have high QAE-pure bounds, and those states can be efficiently compressed using the standard QAE scheme in \cite{ma2023compression}. This suggests that the introduction of quantum mutual information may hinder the performance of low-rank states. 

To maintain good results, it is required to decay the ratio of QMI (i.e., to decrease the value of $\rm w$ in the general function $\Phi(\rm{w})$. This is particularly important for highly pure states. To demonstrate this, we provide the training performance of QAE-qmi with $\rm w=0.5$ and $\rm w=0.99$ in  Fig.~\ref{fig:highpure-2qubit} for 2-qubit states and Fig.~\ref{fig:highpure-4qubit} for 4-qubit states. For states with high purities, training with $\Phi=0.5(J_{\rm{pure}}+J_{\rm{qmi}})$ might lead to fluctuation of decoding fidelity, which can be avoided by reducing the ratio of $J_{\rm{qmi}}$.

The results under the reference scheme $p_r=p_r^{\rm guess}$ are summarized in Fig. \ref{fig:ratio_guess}. 4-qubit states with high $\beta$ are sensitive to the value of $\rm w $ since the performance of $p_r=0.5$ is worse than $\rm w =1$ and $\rm w =0.99$. We further compare the performance of QAE-qmi for the mixture of maximally mixed states and the pure states with results summarized in Fig. \ref{fig:ratio_haar}. The drop of decoding fidelity with $p_0$ approaching one reveals that  $\Phi=0.5(J_{\rm pure}+J_{\rm qmi})$ hinders the decoding fidelity for states with high $p_0$. For such states, increasing the ratio of $J_{\rm pure}$ (i.e., increasing the value of $\rm{w}$) is useful.

\section{Supplementary Discussion 5: Experiments on IBM quantum devices}

Given a state $\rho_K$ of a quantum system $K$, it is possible to introduce another system $R$, and define a pure state $|\psi\rangle_{KR}$ for the joint system $KR$ such that $\rho_K=\textup{Tr}_R(|\psi\rangle \langle \psi|_{KR})$. Suppose $\rho_K$ has orthonormal decomposition $\rho_K=\sum_i p_i |i^K \rangle \langle i^K|$. To purify $\rho_K$, a system $R$ with the same state space as system $K$ with orthonormal basis states $|i^R\rangle$ is introduced. Then, we can define a pure state for the combined system as
$|\psi\rangle_{KR} = \sum_i \sqrt{p_i}|i^K\rangle|  i^R\rangle.$
Its reduced state for system $K$ reads
$$
\begin{aligned}
	\textup{Tr}_R(|\psi\rangle \langle \psi|_{KR}  )=&\sum_{ij} \sqrt{p_ip_j}|i^K\rangle \langle j^K| \textup{Tr}(|i^R\rangle \langle j^R|) \\ =&\sum_{ij}\sqrt{p_ip_j}|i^K\rangle \langle j^K| \delta_{ij}=\rho_K.
\end{aligned}
$$
In this work, we set the orthonormal basis states $|i^R\rangle$ as the computational basis  $|0\rangle,|1\rangle,...$ to generate the associated pure states in the space of $KR$. 

When deploying a unitary transformation on quantum devices, we need to obtain the equivalent of quantum gates that realize the desired unitary transformation. In particular, we utilize the embed function \emph{Initialization} in Qiskit to obtain a sequence of quantum gates that can prepare the target states (i.e., (a) and (c) in Fig. 9 in the main text). Also, we utilize the embed function \emph{Decomposition} in Qiskit to obtain a sequence of gates that achieve the target transformation $U_e$ and $U_d=U_e^{\dagger}$ \cite{shende2005synthesis} (i.e., (b) and (d) in Fig. 9 in the main text).

\begin{figure*}[htb]
	\centering
	\includegraphics[width=0.9\linewidth]{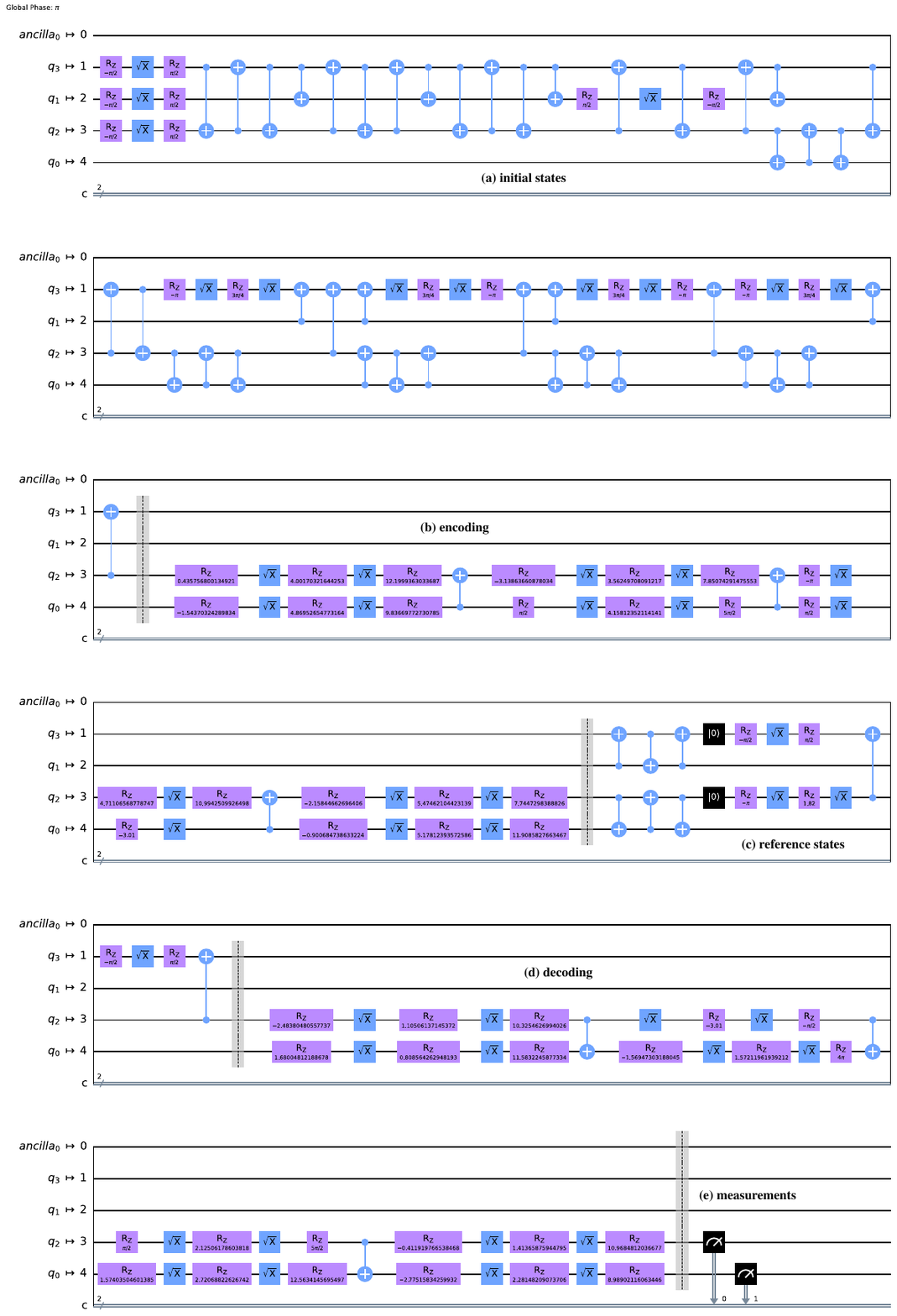}
	\caption{An example of transpiled circuits on \emph{ibmq$\_$qasm$\_$simulator} for compressing a 2-qubit Werner state with $\alpha=0$. The vertical barriers separate different areas of the proposed method, including (a) preparing the initial state, (b) performing the encoding operation, (c), preparing the reference state, (d) performing the decoding operation, (e) performing measurements for quantum state tomography.}
	\label{fig:simulatorcircuit}
\end{figure*}

Here, we provide an example of compressing a 2-qubit Werner state with $\alpha=0$ on  \emph{ibmq$\_$qasm$\_$simulator} in  Fig. \ref{fig:simulatorcircuit}. $U_e$ is decomposed into three 2-qubit CNOT gates and eight 1-qubit $U_3$ gates, where each $U_3$ gate is represented as 
$U_3(\tau, \phi, \lambda)=\left(\begin{array}{cc}
	\cos \left(\frac{\tau}{2}\right) & -e^{\rm{i} \lambda} \sin \left(\frac{\tau}{2}\right) \\
	e^{\rm{i} \phi} \sin \left(\frac{\tau}{2}\right) & e^{\rm{i}(\phi+\lambda)} \cos \left(\frac{\tau}{2}\right)
\end{array}\right).
$

\begin{figure*}[htb]
	\centering
	\includegraphics[width=0.9\linewidth]{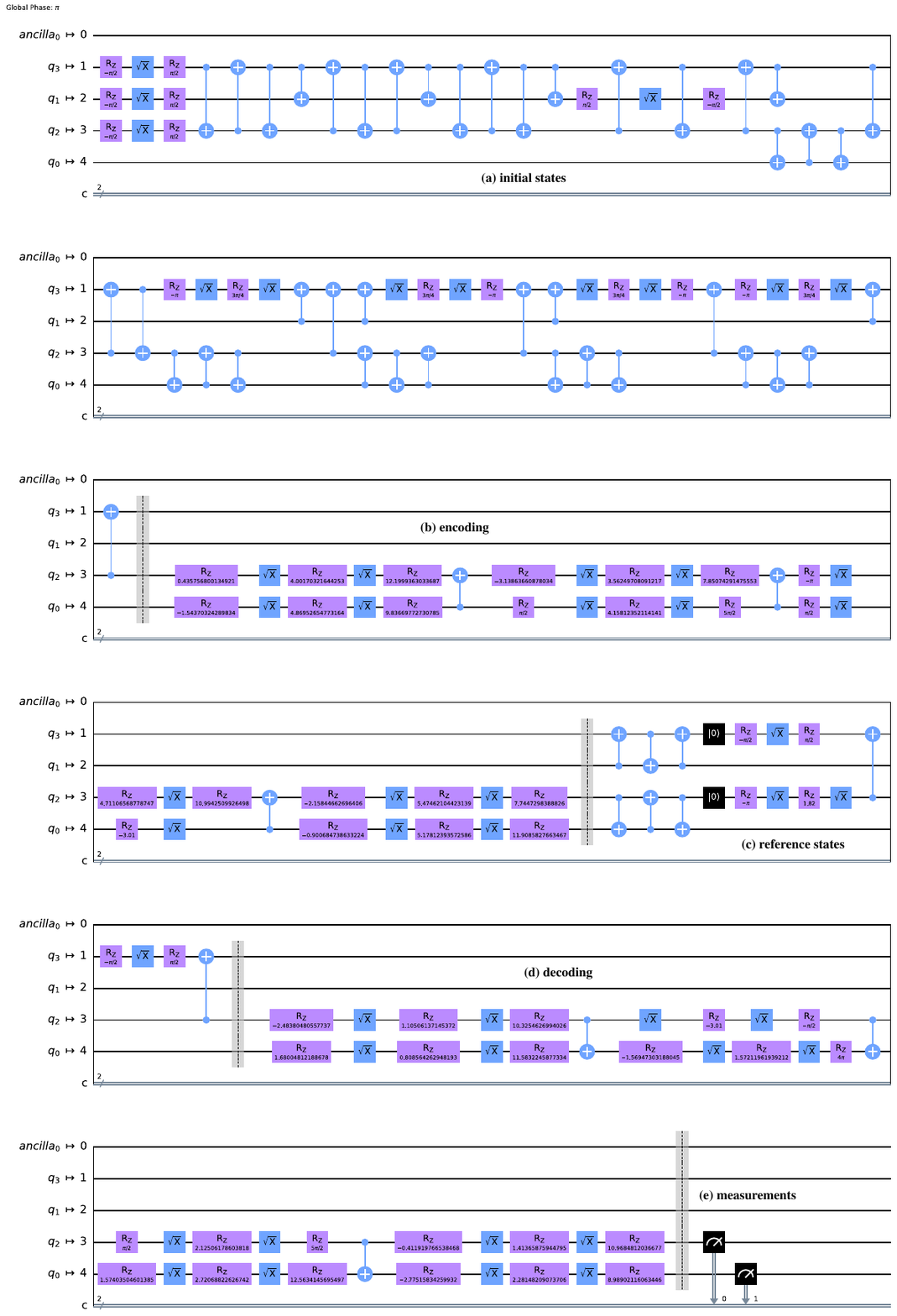}
	\caption{An example of transpiled circuits on \emph{ibmq$\_$quito} for compressing a 2-qubit Werner state with $\alpha=0$. The vertical barriers separate different stages of the proposed method, including (a) preparing the initial state, (b) performing the encoding operation, (c) preparing the reference state, (d) performing the decoding operation, (e) performing measurements for quantum state tomography. }
	\label{fig:quitocircuit}
\end{figure*}

When deploying a sequence of gates on a real quantum computer, some gates can be further decomposed and transpiled into basic gates that are available on the specific quantum device. The transpiled circuits on  \emph{ibmq$\_$quito} for compressing a 2-qubit Werner state with $\alpha=0$ are summarized in  Fig. \ref{fig:quitocircuit}, where only four qubits are utilized among the five available qubits. 

\begin{figure*}[htb]
	\centering
	\includegraphics[width=0.9\linewidth]{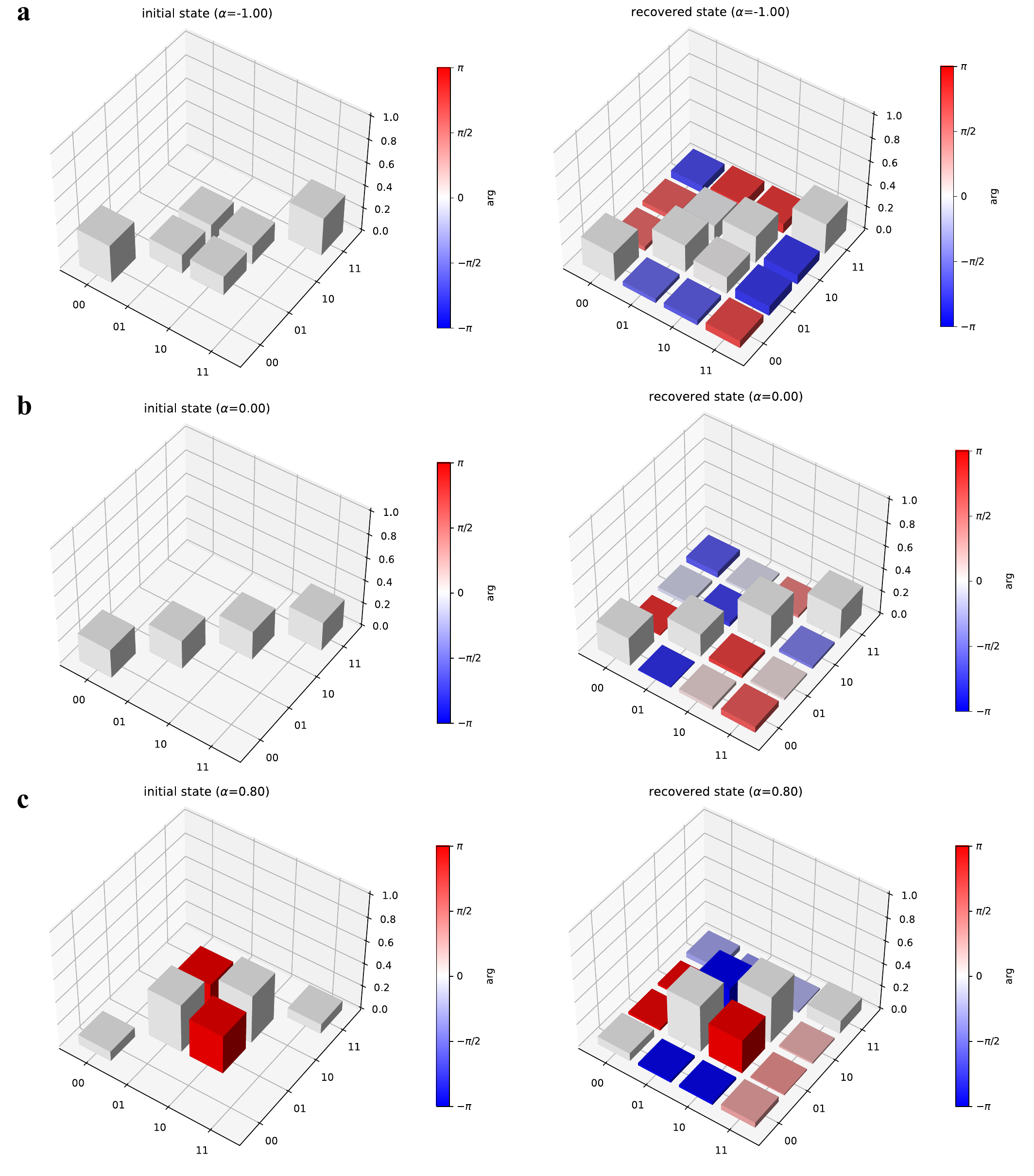}
	\caption{Density matrices of the initial states and the recovered states obtained from \emph{ibmq$\_$quito} for 2-qubit Werner states. The provided figure illustrates three instances of compressing Werner states with different parameters: \textbf{a} for $\alpha=-1.00$, \textbf{b} for $\alpha=0.00$, and \textbf{c} for $\alpha=0.80$. Each 2-qubit state is represented by a $4\times4$ complex matrix, where each element (typically a complex number) is visually depicted as a cylinder. The height of the cylinder represents the amplitude, and the color corresponds to the phase. The density matrices of the recovered states (in the second column) closely resemble the density matrices of the initial states (in the first column). This demonstrates an effective compression of 2-qubit Werner states, preserving essential characteristics in the recovered states. }
	\label{fig:ibmdensitymatrix}
\end{figure*}

\end{document}